\documentclass[bibtotoc,liststotoc,BCOR=5mm,DIV=12, oneside]{scrartcl}

\usepackage[a4paper , lmargin = {3.0cm} , rmargin = {3.0cm} , tmargin = {2.5cm} , bmargin = {4.6cm} ]{geometry} 
\usepackage{adjustbox} 
\usepackage{amsmath}
\usepackage{amssymb}
\usepackage{array}
	\newcolumntype{L}[1]{>{\raggedright\let\newline\\\arraybackslash\hspace{0pt}}m{#1}}
	\newcolumntype{C}[1]{>{\centering\arraybackslash\hspace{0pt}}p{#1}}
\usepackage[english]{babel}
\usepackage{booktabs} 
\usepackage[skip=2pt,font=footnotesize]{caption} 
\usepackage{color}
\usepackage{comment}
\usepackage{csvsimple}
\usepackage{doi} 
\usepackage{eurosym}
\usepackage{graphicx} 
\usepackage{hyperref}
\usepackage{hyphenat} 
\usepackage{lipsum}
\usepackage{listings}
\usepackage{longtable}
\usepackage{makecell} 
\usepackage{multicol} 
\usepackage{multirow} 
\usepackage[round, sort&compress, authoryear]{natbib}
\usepackage{parskip}
\usepackage{scrpage2}
\usepackage{subfig}	
\usepackage{tablefootnote} 
\usepackage{tabularx}
\usepackage[flushleft]{threeparttable}
\usepackage{url}    
\usepackage{xcolor}

\hypersetup{
    colorlinks,
    linkcolor={red!50!black},
    citecolor={blue!50!black},
    urlcolor={blue!80!black}
}  
 
\usepackage[tocgraduated]{tocstyle}
\usetocstyle{standard}

\renewcommand{\baselinestretch}{1.3}

\usepackage[utf8]{inputenc}
\usepackage[T1]{fontenc}

\defcitealias{eeg_in_zahlen_2018}{BMWi, 2018}
\defcitealias{etri_2014}{JRC (2014)}
\defcitealias{statistisches_bundesamt_destatis_stromabsatz_2018}{Destatis, 2018}

\begin{document}


\thispagestyle{empty}
\begin{center}

{\LARGE \textbf{Prosumage of solar electricity: tariff design, capacity investments, and power system effects}}\\

\vspace{10pt}

Claudia Günther$^{1}$, Wolf-Peter Schill$^{1}$, Alexander Zerrahn$^{1,}$*

\vspace{10pt}

\scriptsize{\textit{$^{1}$German Institute for Economic Research (DIW Berlin), Mohrenstr. 58, 10117 Berlin, Germany}} \\
\end{center}

\vspace{25pt}

\normalsize
\textbf{Abstract}: 
We analyze how tariff design incentivizes households to invest in residential photovoltaic and battery systems, and explore selected power sector effects. To this end, we apply an open-source power system model featuring prosumage agents to German 2030 scenarios. Results show that lower feed-in tariffs substantially reduce investments in photovoltaics, yet optimal battery sizing and self-generation are relatively robust. With increasing fixed parts of retail tariffs, optimal battery capacities and self-generation are smaller, and households contribute more to non-energy power sector costs. When choosing tariff designs, policy makers should not aim to (dis-)incentivize prosumage as such, but balance effects on renewable capacity expansion and system cost contribution.

\vfill
\footnotesize
$^*$ \textit{Corresponding author, azerrahn@diw.de} \\
We thank the participants of the Strommarkttreffen seminar at DIW Berlin on 10 May 2019 for valuable feedback on an earlier draft. This work was carried out within the START project, supported with a research grant by the Federal Ministry of Education and Research (BMBF) (grant number 03EK3046E). \\


\section{Introduction}\label{sec:intro}
Decarbonizing energy supply leads to a substantial transformation of the power sector. In many countries, increasing shares of renewable electricity are generated by small-scale distributed solar photovoltaic (PV) plants  \citep{iea_market_2018}. Driven by regulation and dedicated support schemes, private households account for a substantial share of PV installations. In particular, partial self-supply with PV becomes more attractive. Households' self-generation shares can further increase by the advent of PV-plus-battery systems. Batteries have recently experienced a substantial drop in costs; a trend that is expected to continue in the future \citep{schmidt_future_2017}. In this respect, the term prosumage emerged to describe the activity of a household that generates its own PV electricity, enhanced by battery storage, while still being connected to the grid \citep{hirschhausen_2017}.

We investigate how the design of retail and feed-in tariffs (FIT) affects household decisions to invest in PV and battery storage systems. We also explore impacts on the power sector in terms of renewable energy capacities, peak PV feed-in, and the contribution of households to non-energy power system costs, that is, costs for renewable support schemes or the electricity grid infrastructure. To this end, we first provide the intuition how household incentives are shaped by retail tariffs, feed-in tariffs, and respective investment cost for PV and batteries. In a second step, we numerically explore these incentives and their effects in a computational equilibrium model applied to a German 2030 setting. In doing so, we take interactions between households' decisions and the wholesale power market into account.

Central results show that lower feed-in tariffs substantially reduce PV investments. Yet effects on battery capacity and PV self-generation are less pronounced. Higher fixed parts and lower volumetric components in retail tariffs lead to lower optimal battery capacities and self-generation. In turn, households contribute more to non-energy system costs. We further find that limiting peak feed-in, which can relief distribution grids, is possible without substantially distorting households' incentives. 

While previous research features a number of analyses on the economic viability of prosumage for specific households, only few contributions consider feedback of prosumage on the power sector. Yet these are largely silent on household behavior and the impact of the regulatory setting. Combining the household, power sector, and regulatory policy perspectives, this paper aims to fill a gap in the literature. It contributes to the academic and policy debates on retail tariff design, increasing deployment of decentral PV and storage systems, and contribution to non-energy power sector costs.

The remainder of this paper is structured as follows. Section~\ref{sec:litreview} reviews relevant literature on prosumage. Section~\ref{sec:prosumage} introduces a conceptual framework and provides some intuition. Based on this, we develop a formal equilibrium model in section~\ref{sec:model}. Section~\ref{sec:results} presents the numerical results; section~\ref{sec:limitations} discusses limitations of our approach and outlines avenues for future research. Section~\ref{sec:conclusion} concludes.


\section{Literature Review}\label{sec:litreview}
This paper contributes to three overlapping strands of the literature on the economics of prosumage: the household perspective, the power sector perspective, and the policy perspective.

First, concerning the household perspective, a range of publications analyzes the economic viability of prosumage systems. For Germany~--~a country with favorable market conditions~--~\citet{hoppmann_economic_2014} were among the first to argue that falling battery costs will spur the prosumage segment. \citet{kaschub_solar_2016} come to a similar conclusion: according to their analysis, prosumage will be economical for households in the short run even in the absence of subsidies, driven by self-consumption. Using self-generated electricity for electric vehicles would support this trend. \citet{dietrich_what_2018} also conclude on the profitability of PV-plus-battery systems in the near future, and highlight that economies of scale would incentivize larger installations. A comparative study for Germany and Ireland derives analogous results \citep{bertsch_what_2017}. Studies on the viability of residential prosumage also exist for electricity markets in other countries like Australia~\citep{muenzel_pv_2015, say_coming_2018}, Brazil~\citep{gomes_technical-economic_2018},  France~\citep{yu_prospective_2018}, Italy~\citep{cucchiella_photovoltaic_2016}, Spain~\citep{prol_photovoltaic_2017, solano_impact_2018}, the United Kingdom~\citep{green_prosumage_2017}, and the United States~\citep{khalilpour_leaving_2015, say_coming_2018, tervo_economic_2018}. 

Second, concerning the power sector perspective, prosumage contrasts with the traditional supply- and demand-side division. A widespread adoption of residential PV-plus-battery systems affects both prosumage households and other electricity consumers as well as power generators. This could have broad technical, socio-economic, and political repercussions, as discussed by~\citet{agnew_effect_2015}, \citet{schill_prosumage_2017}, and~\citet{schill_2019}. More specifically, residential PV-plus-battery systems provide low-carbon energy and can thus help to achieve climate targets. However, depending on prosumagers' price signals and objectives, economic inefficiencies may arise. These relate to sub-optimal investment in the long run, for instance redundant storage infrastructure, and sub-optimal dispatch in the short run, for instance a modest contribution to system peak shaving or valley filling \citep{green_prosumage_2017, schill_prosumage_2017}. 

Besides investments and dispatch, \citet{marwitz_techno-economic_2018}, \citet{moshovel_analysis_2015}, and~\citet{neetzow_2019} discuss whether and under which regulatory circumstances prosumage may require expanding the electricity distribution grid infrastructure. In contrast, \citet{young_2019} conclude for an Australian region that benefits from reduced peak grid demand may outweigh foregone revenues from self-consumption for network operators.  

Third, concerning the regulatory policy perspective, an increasing number of studies addresses the question how to price prosumagers' consumption and generation. Proposed policies can be broadly divided into a revision of retail rate structures, remuneration schemes for decentral generation, and other policy measures. For the German context, \citet{ossenbrink_how_2017} investigates the implications of feed-in and retail tariff schemes for residential PV systems without batteries. Based on the ratios of the levelized cost of electricity (LCOE) of PV to the feed-in and retail tariffs, respectively, he derives conditions under which households act as pure consumers or engage in prosuming activities. However, storage is excluded from the analysis. More recently, \citet{thomsen_jessica_how_2019} assess how the tariff design affects profitability and operation of small-scale PV-plus-battery systems in Germany, yet assuming fixed prosumage PV and battery capacities and not analyzing investment incentives.

Broadening the scope, tariff design triggers re-distributive effects in the context of prosumage. Generally, volumetric energy charges lead to a burden shift from prosumage households to pure consumers \citep{roulot_2018}. Since prosumage households have a lower grid energy consumption (\textit{load defection}), they pay fewer network fees and other charges although they still enjoy all grid services~\citep{simshauser_distribution_2016}. A growing prosumage segment can induce distributive justice concerns at the consumer level and cost recovery issues for utility operators \citep{hinz_regional_2018, kubli_squaring_2018, roulot_2018, schittekatte_future-proof_2018}, leading to what has been referred to as \textit{death spiral} in the most extreme case~\citep{costello_2014, laws_utility_2017}: ever fewer customers must pay ever higher grid charges which, in turn, incentivizes further load defection. However, given that this would require major uptake of residential PV-plus-battery systems, such scenario is rather unlikely to occur in many developed countries' power sectors~\citep{darghouth_net_2016, laws_utility_2017}.\footnote{A related debate discusses the metering design in case of residential photovoltaic (plus battery) systems. Specifically, under net metering consumers' PV electricity grid feed-in and and grid-consumption are offset against each other over a longer time horizon~\citep{hughes_compensating_2006}. This was shown to substantially raise incentives for residential standalone PV investments in different settings \citep[e.g.,][]{eid_economic_2014, picciariello_electricity_2015, darghouth_net_2016}, fuelling discussions on distributional justice.}

In conclusion, the literature features individual analyses that provide an in-depth treatment of specific households, yet mostly do not take interactions between prosumagers and the power sector into account. Power sector studies mostly neither incorporate prosumage households' incentives nor the tariff design. Regulatory studies generally lack a numerical underpinning. We aim to contribute to all three perspectives and provide an analysis of tariff design for prosumage, which also includes PV and storage investment incentives of households. While we focus on the German context, results are of interest also for other markets where feed-in tariffs and high volumetric retail tariffs drive solar prosumage.

    
\section{Residential prosumage: definitions and intuition}\label{sec:prosumage}


\subsection{Definitions}\label{subsec:prosumage_definition}
\begin{figure}[hbt!]
\centering
\footnotesize
\includegraphics[height=9.0cm]{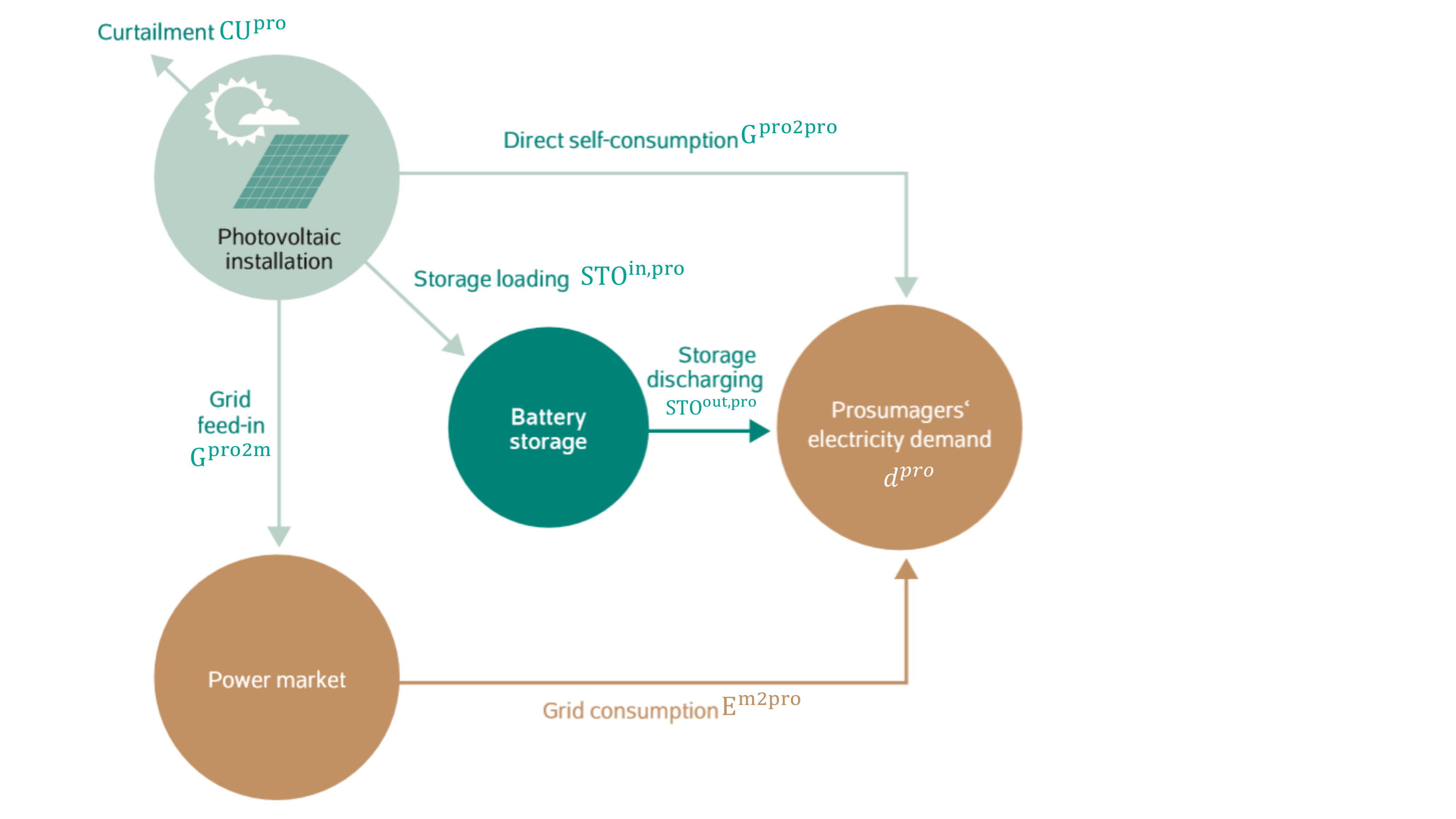}
\caption{Schematic representation of energy flows of a prosumage household. Own illustration, based on \citet{schill_2019}}
\label{fig:scheme_prosumage}
\end{figure}

We define a residential prosumager as a grid-connected household with a PV panel and a battery \citep{schill_prosumage_2017}. Figure~\ref{fig:scheme_prosumage} presents a stylized representation. The household generates electricity~($G^{PV}$) that it consumes at times~($G^{pro2pro}$), feeds into the grid at other times~($G^{pro2m}$), potentially curtails~($CU^{pro}$), or stores in the battery~($STO^{in,pro}$) for future consumption~($STO^{out,pro}$). Still, the household may consume electricity from the grid, i.e.~the market, at any point in time~($E^{m2pro}$). For clarity, we assume that the battery can only be used for deferring self-consumption.\footnote{In principle, other uses are conceivable like smoothing grid consumption and feed-in or providing flexibility by storing in and out grid electricity. \citet{schill_prosumage_2017} show that such additional battery use can lower total power system costs.}

There are two standard metrics that describe the dependency of prosumage households on power provision from the grid: the rate of self-consumption and the autarky rate \citep{luthander_photovoltaic_2015, weniger_sizing_2014}. The rate of self-consumption~$SC$ is the fraction of electricity generated on-site that is either directly consumed or stored in the battery for future self-consumption. 
\vspace{0.5cm}
\begin{subequations}
\begin{equation}
SC := \frac{G_{}^{pro2pro} + STO_{}^{in,pro}}{G^{PV}} 
\end{equation}

The lower the rate of self-consumption~$SC$, the higher the revenues generated from feeding energy into the grid. For example, a self-consumption rate of~$40$\% implies that~$60$\% of the generated electricity receive a remuneration, for example the feed-in tariff. This explains why relying solely on levelized cost of electricity (LCOE) to determine the profitability of a PV-plus-battery system for a single household is inadequate. 

The autarky rate~$A$, also referred to as rate of self-generation or rate of self-sufficiency, is the share of household electricity demand covered by generation from the PV-plus-battery system. This includes directly consumed energy and energy discharged from the battery storage. 
\vspace{0.5cm}
\begin{equation}
A := \frac{G^{pro2pro} + STO^{out,pro}}{d^{pro}} 
\end{equation}
\end{subequations}

The higher a household's the autarky rate~$A$, the less it pays for energy consumed from the grid. Both the autarky and self-consumption rates are defined over a specified time interval, usually a year.
\vspace{0.5cm}


\subsection{Intuition: Incentives for prosumage}\label{subsec:prosumage_households}
\begin{figure}[hbt!] 
\centering
\footnotesize
\includegraphics[height=7.0cm]{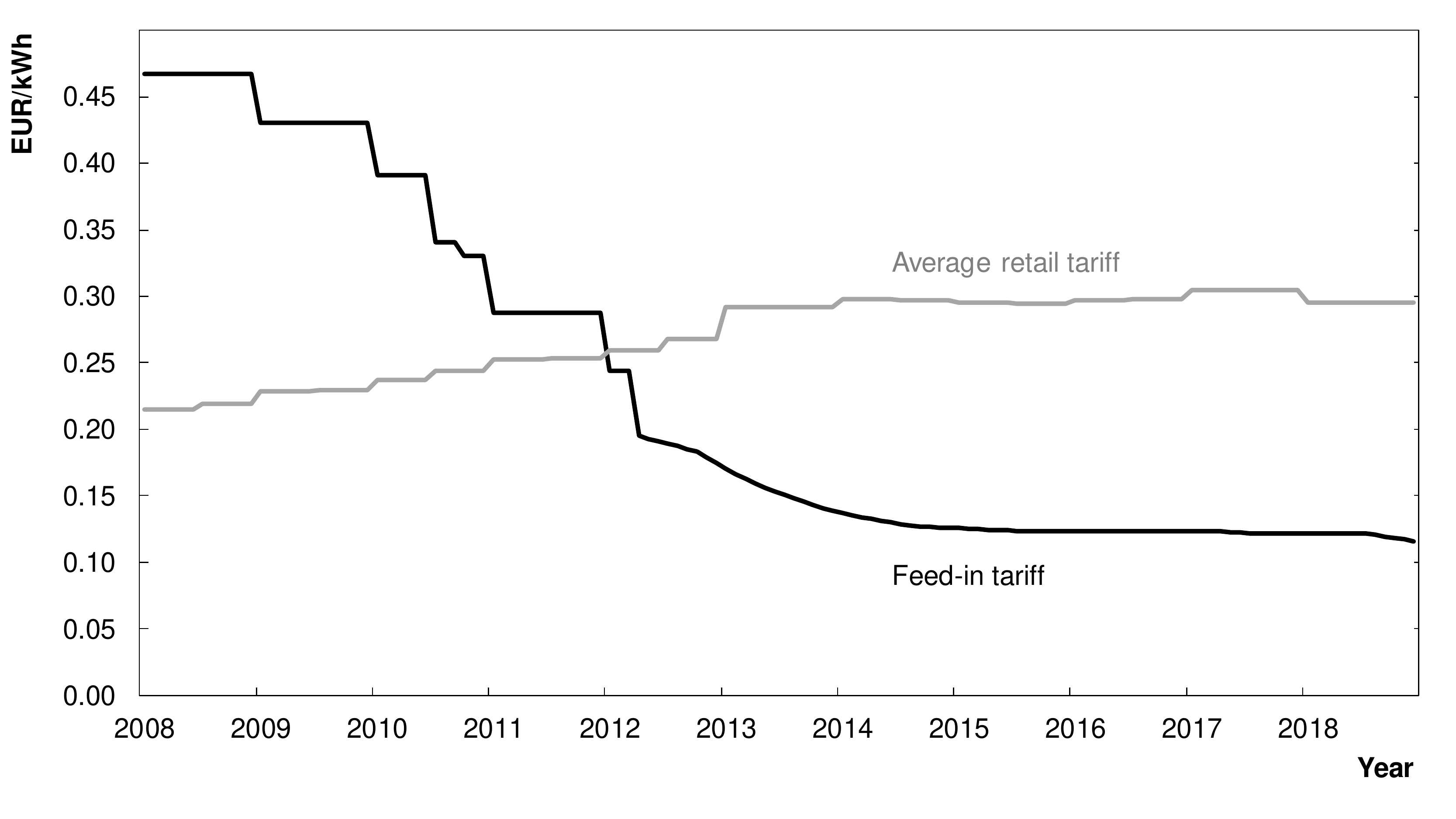} 
\caption{Development of the feed-in tariff for small-scale PV and the average residential electricity retail tariff in Germany. Source: Own illustration with data from~\citet{bnetza_eeg-registerdaten_2018}.}
\label{fig:timeseries_FIT_retail}
\end{figure}

For a rational household, economic incentives for prosumage are largely driven by investment costs for PV and storage systems as well as grid consumption and feed-in tariffs.\footnote{See~\citet{schill_2019} for an overview of other motivations for prosumage and~\citet{gautier_2019} for an empirical investigation.} Under the current German setting, households face time-invariant, volumetric rates for grid consumption and PV grid feed-in, the retail tariff and the feed-in tariff, respectively. In 2008, the average German residential retail tariff for electricity was~\text{0.21 EUR/kWh}. A decade later, in 2018, it had increased to about~$0.30$~EUR/kWh, of which~$0.23$ EUR/kWh accrued for non-energy components like taxes, network charges and renewable support payments; a share of~$80$\%. Importantly, self-consumption from small-scale PV systems below~$10$~kW is exempt from any such cost components. At the same time, the cost degression of PV systems has led to a substantial reduction of feed-in tariffs, which have decreased from more than~\text{0.46 EUR/kWh} in~2008 to less than~\text{0.12 EUR/kWh} in~2018. Figure~\ref{fig:timeseries_FIT_retail} illustrates these developments over the last decade.
\vspace{0.5cm}

In principle, if the retail tariff exceeds the FIT, self-consumption of PV electricity becomes economical. Yet this \textit{socket parity} \citep{bazilian_2013} does not necessarily imply economic viability of prosumage because residential demand and PV supply may only partly coincide in time. Savings from further substituting grid consumption with self-consumption must over-compensate the necessary investments into a battery. 

\begin{figure}[hbt!]
\centering
\footnotesize
\includegraphics[height=7.4cm]{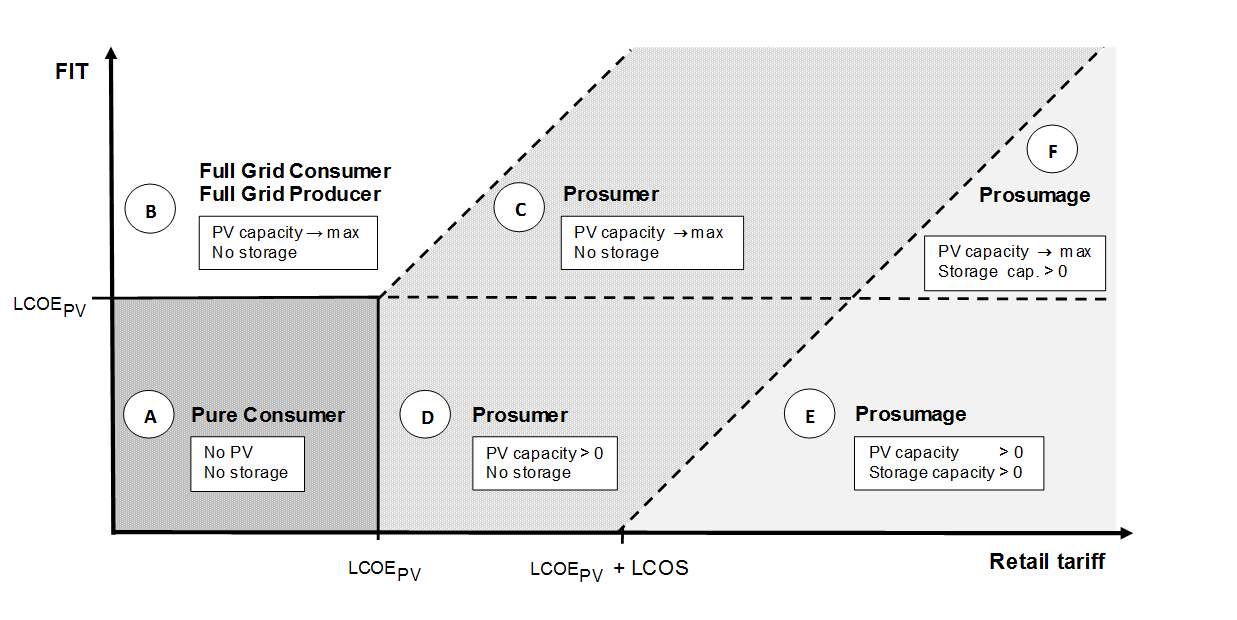}
\caption{Illustration of incentives for investments in residential PV and battery systems as a function of feed-in tariffs (FIT), retail tariffs, and the LCOE of PV. Source: Own illustration based on \cite{ossenbrink_how_2017}.}
\label{fig:ossenbrink}
\end{figure}

To this end, figure~\ref{fig:ossenbrink} provides a more comprehensive illustration of incentives for investments in residential PV and battery systems. It extends an analysis by~\citet{ossenbrink_how_2017} by adding the storage dimension. In the following, we go through the figure's areas~$A$~to~$F$, using the German situation as an example. The lower left area~$A$ refers to a situation where the LCOE of a decentral PV installation exceed both the retail and feed-in tariffs. Hence, there is no financial incentive for households to invest in PV or battery systems (``Pure consumer''). This was the situation in Germany before a FIT was introduced in the year 2000.

The 45-degree line starting at the upper right corner of area~$A$ marks the points where the FIT and retail price are equal. In the upper left area~$B$, the FIT exceeds the levelized costs of PV, and the FIT is also higher than the retail tariff. This characterizes the market situation in Germany before 2012. Households have an incentive to feed all generated PV electricity into the grid and satisfy their demand with grid consumption.\footnote{For completeness, a provision granted a bonus on self-consumption in Germany between~2009 and~2012. We do not depict or illustrate this particular regulation.} This means they are full grid producers and consumers. Moreover, since the PV system can generate positive revenues, households are incentivized to invest in a PV system that is as large as possible. Yet there is no incentive to install battery storage since it is more attractive to generate revenues from the FIT than to avoid consuming from the grid.

When the retail tariff exceeds both the LCOE of PV and the FIT, self-consumption becomes attractive (``Prosumer'' areas~$C$~and~$D$). Consider area~$C$ first. The FIT exceeds the levelized costs for PV and creates an incentive to install the maximum PV capacity. At the same time, the retail tariff is higher than the FIT, meaning that it is attractive to substitute as much grid consumption as possible with on-site generation. In area $D$, the  FIT is not high enough to cover the LCOE of PV. In this situation, households no longer have an incentive to install the maximum possible PV capacity. The optimal PV capacity trades off the costs for the PV system with the revenues collected through the FIT and the expenditure on grid consumption saved through self-consumption.
\vspace{0.5cm}

In both areas~$E$ and~$F$, installing a battery storage is more profitable than a PV-stand-alone system. For this to be the case, the retail tariff must not only be high relative to the LCOE of PV, but also relative to the FIT. Since the battery is used to offset grid consumption with self-generated PV energy, which would otherwise be fed to the grid, the difference between the two tariffs must cover the levelized costs of storage (LCOS). The higher the storage costs, the further areas~$E$ and~$F$ are shifted to the right. Similarly, battery capacity deployed by households increases with this difference since it increases incentives for offsetting grid consumption. As before, these two prosumage situations differ with respect to the deployed PV capacity. If the FIT is above the LCOE of PV, in area~$F$, households deploy the maximum PV capacity, while the optimal PV size is smaller in~$E$. Households are incentivized to maximize their self-consumption in both cases. The market situations depicted in $E$ and $F$ have not yet been reached in Germany by 2019 since storage costs are, still, too high. 


\section{Model, data, and scenarios}\label{sec:model}
In the following, we illustrate the incentive structure by means of a numerical model applied to German 2030 scenarios. Beyond prosumage household decisions, we investigate selected effects on the power sector.


\subsection{Model}\label{subsec:model_setup}
The model features a prosumage household agent and a benevolent power sector operator. We assume that both agents have perfect foresight. Each solves a linear cost minimization problem such that the Karush-Kuhn-Tucker (KKT) conditions are both necessary and sufficient for global optimality of the solution. Throughout the mathematical exposition, capital letters denote variables and lower-case letters parameters. 

The prosumage agent minimizes her annual electricity expenditure~$Z^{pro}$ by deciding on optimal prosumage system investment and dispatch \eqref{eq:obj_pro}. Retail costs for grid electricity consumption consist of a fixed component and a volumetric component for hourly electricity consumption from the grid~$E^{m2pro}_h$. The volumetric component comprises a price for energy~$t^{ener}_h$, which may vary from hour to hour, and a time-invariant part~$t^{other}$, which represents non-energy charges like taxes, network fees or renewable support surcharges. Non-energy charges may also be raised by a fixed annual component~$t^{fix}$. Depending on the scenario, cost components may be zero or positive. To avoid costly grid consumption, the household can invest in PV capacity~$N_{pv}^{pro}$ as well as lithium-ion battery energy and power capacities, $N_{sto}^{pro,E}$ and~$N_{sto}^{pro,P}$, respectively. Annualized investment costs are factored in via~$c^{inv}$ and the annual fixed costs via~$c^{fix}$. Investments into PV, battery power, and battery energy capacities are mutually independent. Prosumage households can also lower their annual electricity bill by generating positive revenues through selling energy~$G^{pro2m}_h$ to the grid at a, potentially time-varying, price of~$t^{prod}_h$.  
\vspace{0.2cm}
\begin{subequations}\label{eq:household_problem}
\begin{align}
\mbox{ min} \hspace{0.1cm} Z^{pro} = &\sum_{h}\left[E^{m2pro}_h*\left(t^{ener}_h +  t^{other}\right)\right] + t^{fix}\label{eq:obj_pro}  \nonumber \\
- &\sum_{h}\left(G^{pro2m}_h * t^{prod}_h\right) \nonumber \\
+  & N_{pv}^{pro} \left(c^{inv}_{pv} + c^{fix}_{pv}\right)   \nonumber \\
+ & N_{sto}^{pro,E}(c^{inv,E}_{sto} + \textstyle\frac{1}{2} c^{fix}_{sto})     + 
N_{sto}^{pro,P}\left(c^{inv,P}_{sto} + \textstyle\frac{1}{2} c^{fix}_{sto}\right)
\end{align}

Households' inelastic electricity demand~$d^{pro}_h$ must be satisfied in each hour, either with directly consumed PV generation~$G^{pro2pro}_h$, through energy discharged from storage~$STO^{out,pro}_h$ or with grid consumption~\eqref{eq:ener_balance}. For each constraint, the respective Lagrange multiplier, or shadow price, is given in parentheses.
\vspace{0.2cm}
\begin{align}
d^{pro}_h &= G^{pro2pro}_h + STO^{out,pro}_h + E^{m2pro}_h && \forall h\label{eq:ener_balance} &&& (\lambda^{enbal,pro}_{h})
\end{align}

The hourly available PV energy depends on the exogenous capacity factor~$\phi^{avail}_{pv,h} \in \left[0,1\right]$ and the installed capacity. It can be consumed by the household, sold to the market, curtailed~$CU^{pro}_h$ or stored~$STO^{in,pro}_h$~\eqref{eq:pv_use}.
\vspace{0.2cm}
\begin{align}
\phi^{avail}_{pv,h}* N_{pv}^{pro} &= G^{pro2pro}_h + G^{pro2m}_h  + CU^{pro}_h +
STO^{in,pro}_h  && \forall h\label{eq:pv_use} &&& (\lambda^{pv,pro}_{h})  
\end{align}

The stored energy at the end of each hour~$STO^{l,pro}_h$ is equal to the storage level at the end of the previous period, minus the energy discharged in the current period, plus the charged energy, corrected by the battery's roundtrip efficiency~$\eta$~\eqref{eq:sto_l}. In the first period, the household starts with an empty storage~\eqref{eq:sto_l1}. Battery use is subject to the installed power and energy capacities~(\ref{eq:sto_maxlev}--\ref{eq:sto_maxout}).
\vspace{0.2cm}
\begin{align}
STO^{l,pro}_h &=  STO^{l,pro}_{h-1}
+ \textstyle\frac{1+ \eta_{sto}^{pro}}{2}*STO^{in,pro}_h \nonumber\\
& \hspace{2cm} - \textstyle\frac{2}{1+ \eta_{sto}^{pro}}*STO^{out,pro}_h && \forall h > h_1 &&&(\lambda^{sto,pro}_{h}) \label{eq:sto_l}\\\nonumber\\ 
STO^{l,pro}_{h_1} &=  
\textstyle\frac{1+ \eta_{sto}^{pro}}{2}*STO^{in,pro}_{h_1}- \textstyle\frac{2}{1+ \eta_{sto}^{pro}}*STO^{out,pro}_{h_1} && &&&(\lambda^{sto,pro}_{h_1})\label{eq:sto_l1}
\end{align} 
\begin{align}
STO^{l,pro}_h &\leq N_{sto}^{pro,E} && \forall h &&& (\lambda^{stol,pro}_{h}) \label{eq:sto_maxlev} \\ 
STO^{in,pro}_h  &\leq N_{sto}^{pro,P} &&\forall h  &&& (\lambda^{stoin,pro}_{h}) \label{eq:sto_maxin} \\
STO^{out,pro}_h &\leq N_{sto}^{pro,P} &&\forall h  &&& (\lambda^{stoout,pro}_{h}) \label{eq:sto_maxout}
\end{align}

The maximum PV capacity per household is limited to~$m_{pv}$ to reflect space restrictions or regulatory thresholds~\eqref{eq:max_pv}. Appendix~\ref{app:subsec:model_lagrange_hh} gives a full account of the Lagrangian and according KKT first-order optimality conditions.
\vspace{0.2cm}
\begin{align}
N_{pv}^{pro} &\leq m_{pv} && &&& (\lambda^{pvmax,pro}) \label{eq:max_pv}
\end{align}
\end{subequations}

The second agent is a benevolent power sector operator. She minimizes total system costs through optimal dispatch of a given power plant and pumped-hydro storage fleet. This dispatch is equivalent to a competitive market outcome. The market clears in every hour, meaning that total generation must equal total consumption. This includes both the grid demand from the prosumage household and a non-prosumage inelastic demand representing all other consumers. The according dual variable~$\lambda^{enbal}_h$ can be interpreted as the wholesale market price that is passed on to the prosumage household in some scenarios. The power  sector dispatch is based on the open-source model DIETER \citep{dieter_2017}. Appendix~\ref{app:subsec:model_lagrange_sys} lists all equations.
\vspace{0.5cm}
\begin{figure}[hbt!]
\centering
\footnotesize
\includegraphics[height=8cm]{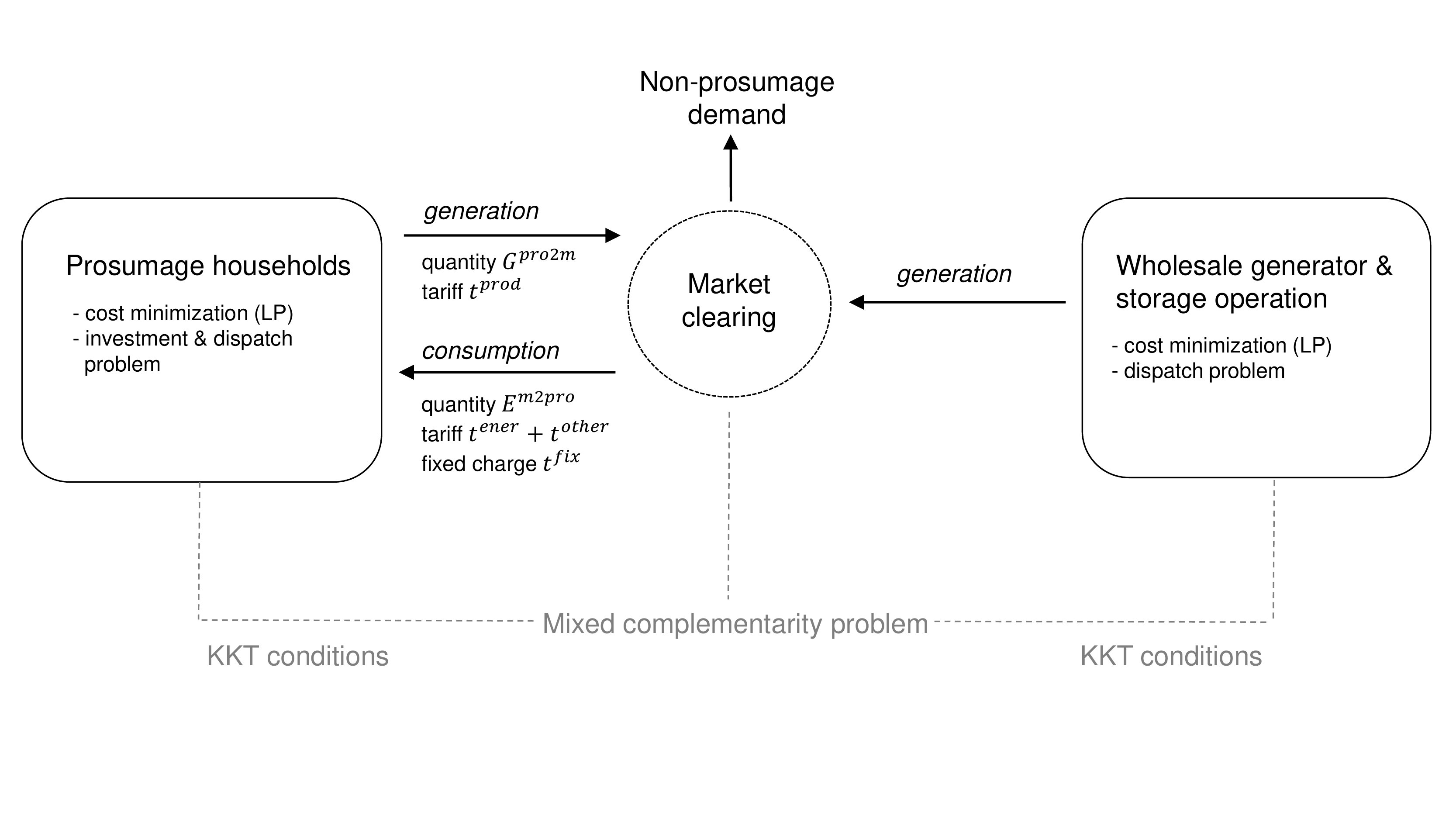}
\caption{Illustration of model set-up as equilibrium problem between prosumage households and the power sector featuring wholesale generators and storage operators.}
\label{fig:mcp_scheme}
\end{figure}

In order to analyze prosumage households and generators jointly in an equilibrium problem, we combine the KKT conditions of the two programs in a mixed complementarity problem (MCP) \citep{facchinei_finite-dimensional_2007}. Figure~\ref{fig:mcp_scheme} illustrates the interaction between the wholesale generators as well as storage operators and the prosumage households. The problem is implemented in GAMS and solved with the PATH solver \citep{dirkse_path_1995}. All model sets, decision variables, and parameters are listed in tables~\ref{tab:app:sets}, \ref{tab:app:variables} and~\ref{tab:app:parameters} in appendix~\ref{app:subsec:model_sets}. The model is solved for every second hour of the year. This allows capturing all the important daily and seasonal demand variability as well as fluctuating generation of wind and solar plants while ensuring tolerable computation times. We provide the model code and input data under  a  permissive  open-source  license  in  a  public repository.\footnote{\url{https://doi.org/10.5281/zenodo.3345784}} 


\subsection{Data} \label{subsec:model_data}
The size of the prosumage segment is set to one million households, in accordance with prosumage growth projections for Germany \citep{bnetza_genehmigung_2018}. This number represents approximately ten percent of all single-family and two-family houses that are potentially suitable for PV-plus-battery systems \citep{prognos_eigenversorgung_2016}. That is, the prosumage agent represents the aggregate of all prosumage households. In line with the current threshold for being exempt from paying the renewable surcharge on self-consumed electricity, we set an upper PV investment limit of~$10$~kW per household. The annual load of each prosumage household is assumed to be~$5$~MWh, the average value of a German single-family household \citepalias{statistisches_bundesamt_destatis_stromabsatz_2018}. 
\vspace{0.5cm}
\begin{table}[htb!]
\caption{Selected parameters for the prosumage segment}\label{tab:prosumage_par}
\centering  
\footnotesize
\begin{threeparttable}
\begin{tabular}{llll}
\midrule
 & Value & Unit & Source \\ 
\midrule
Interest rate  & 4\% & &  Own assumption  \\
VAT & 19\% & &  Own assumption  \\
\midrule
\textbf{Residential photovoltaics} \ &&&  \\
Overnight investment costs & 850 & EUR/kW & \citetalias{etri_2014}\\
Technical lifetime  & 25 & years &  \citetalias{etri_2014} \\
Annual fixed costs \textit{$c^{fix}_{pv}$} & 17 & EUR/kW & \citetalias{etri_2014} \\
Annualized investment costs \textit{$c^{inv}_{pv}$} & 64.75 & EUR/kW & Own calculation \\
Annual full load hours & 1090 & h &  \citet{pfenninger_long-term_2016} \\
\midrule
\textbf{Residential lithium-ion batteries} &&&  \\
Round-trip efficiency \textit{$\eta_{sto}^{pro}$} & 0.92 & & \cite{pape_roadmap_2014} \\
Overnight investment costs power & 140 & EUR/kW & \citetalias{etri_2014}\\
Overnight investment costs in energy & 205 & EUR/kW & \citetalias{etri_2014} \\
Technical lifetime & 15 & years & Own assumption \\
Annual fixed costs \textit{$c^{fix}_{sto}$}& 10 & EUR/kW & Own assumption \\
Annualized investment costs power \textit{$c^{inv,P}_{sto}$} & 14.98 & EUR/kW & Own calculation  \\
Annualized investment costs energy \textit{$c^{inv,E}_{sto}$} & 21.94 & EUR/kWh & Own calculation \\
\bottomrule
\multicolumn{4}{l}{\begin{tabularx}{\linewidth}{X}\textit{Note: Referenced values exclude VAT as stated by the source. Own calculations of annualized costs include VAT.} \end{tabularx}}	\\
\end{tabular}
\end{threeparttable}
\end{table}

Table~\ref{tab:prosumage_par} lists the relevant parameters for the prosumage segment, mainly drawing on the European Energy Technology Reference Indicator projection (ETRI) for 2030~\citepalias{etri_2014}. 
The expected LCOE of small-scale PV for a household range between~$0.07$ and~$0.08$~EUR/kWh, including expenses for value-added tax (VAT). The assumed storage costs imply a substantial reduction compared to current levels in Germany. Though this projection is optimistic regarding storage cost decline, it lies within the expected range for 2030 reported by~\citet{schmidt_future_2017}.

Demand of prosumage households follows the standard load profile of the German Association of Energy and Water Industries \citep{bdew_standardlastprofile_2015} and is scaled up to represent one million households. The hourly PV capacity factor is a capacity-normalized, country-aggregated time series taken from~\citet{pfenninger_long-term_2016} and~\citet{staffell_using_2016}. These sources are based on reanalysis data of the year~2012, which constitutes an average weather year. 

\begin{figure}[hbt!]
\centering
\footnotesize
\includegraphics[height=8.0cm]{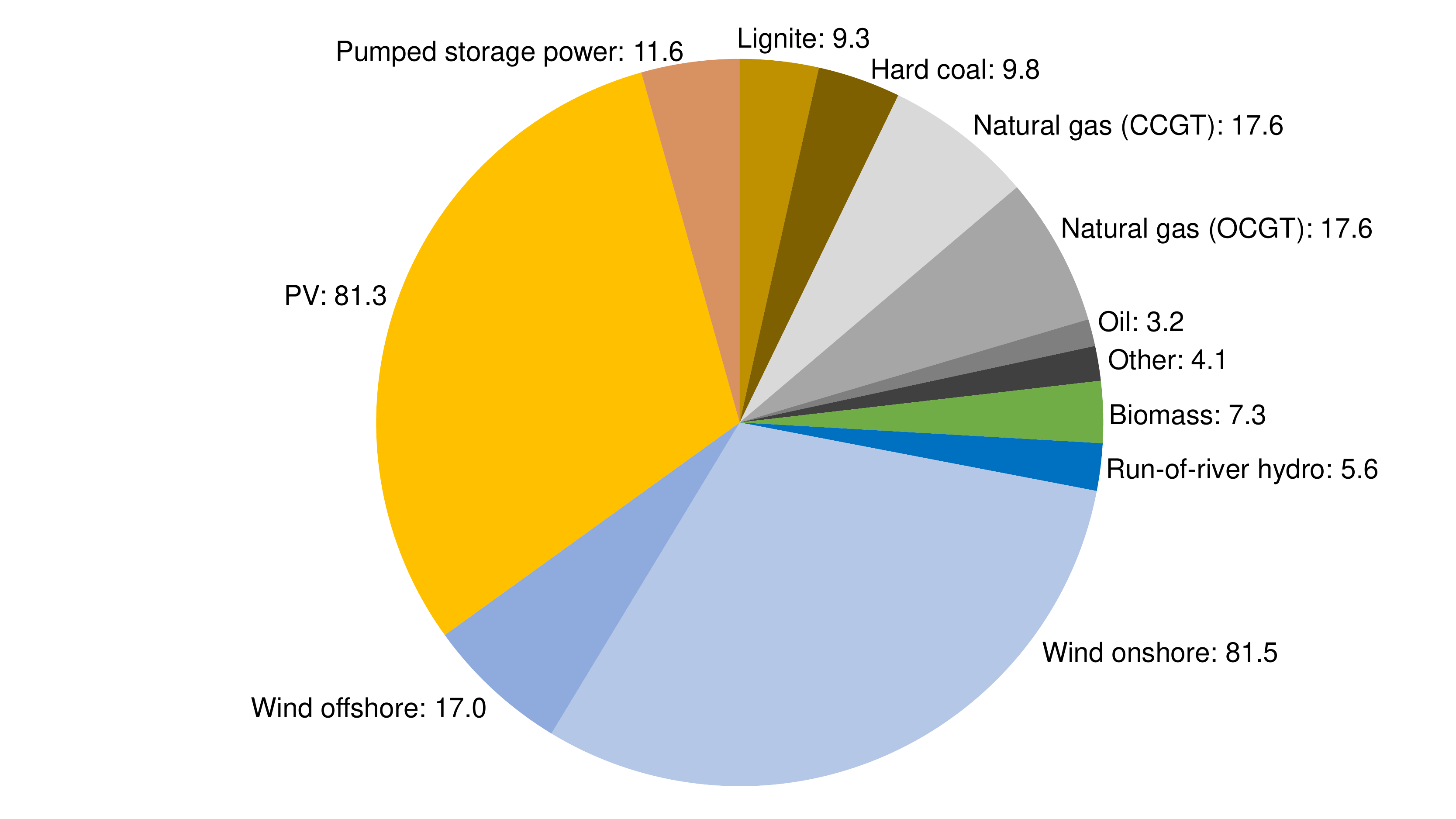}
\caption[]{Assumed power sector generation and storage capacities in~$GW$ based on the middle scenario B of the German Grid Development Plan 2030 \citep{bnetza_genehmigung_2018}.}
\label{fig:NEP_2030}
\end{figure}

The power sector is calibrated in line with official projections for Germany in~2030. Generation and storage capacities shown in figure~\ref{fig:NEP_2030} correspond to the middle scenario B in the latest release of the approved German Grid Development Plan 2030 (NEP 2030) of the Bundesnetzagentur, the German network agency \citep{bnetza_genehmigung_2018}. Costs and technical parameters for operation of power plants and storage are taken from the Grid Development Plan whenever possible and completed with information from~\citet{schroder_current_2013} and~\citet{pape_roadmap_2014}. Table~\ref{tab:app:parameters_system} in appendix~\ref{app:subsec:model_data} lists the complete cost parameters for the power sector dispatch. 

The German load profile for pure consumers follows the Ten-Year Network Development Plan 2030 of the European Network of Transmission System Operators for Electricity~\citep{entsoe_tyndp_2018}. This projection of future demand is supposed to be representative of a normal weather year. We subtract the demand of the prosumage segment from the national demand curve. As for renewable energy sources, hourly capacity factors are taken from~\citet{pfenninger_long-term_2016} and~\citet{staffell_using_2016}.
\vspace{0.5cm}


\subsection{Scenarios}\label{subsec:model_scenarios}
We devise a number of scenarios of the year~2030 that differ with respect to which price signals prosumage households receive (table \ref{tab:scenarios}). Some scenarios also include other, non-price-based policies. 
\vspace{0.5cm}
\begin{table}[h!]
\caption{Scenarios}
\centering\label{tab:scenarios}
\begin{threeparttable}  
\begin{adjustbox}{width=\textwidth}
\begin{tabular}{lccccc}
\toprule   \\
 & \multicolumn{3}{c}{\textbf{Retail tariff components}} & \multicolumn{2}{c}{\textbf{Feed-in tariff components}} \\
 \midrule
 & \textbf{Energy charge} & \textbf{Other charge} & \textbf{Fixed part} &  \textbf{FIT|RTP}  & \textbf{Other policies} \\
 & [EUR/kWh] & [EUR/kWh] & [EUR/year]  & [EUR/kWh]  &   \\
 & $t^{ener}_{(h)}$ & $t^{other}$ & $t^{fix}$ &  $t^{prod}_{(h)}$ &  \\
\midrule
\\ 
\multicolumn{6}{l}{\textbf{a) Scenarios with a purely volumetric retail tariff}}  \\ [0.75ex] 
Retail\_30 FIT\_8 (Baseline) & $0.05$ & 0.25 & - & 0.08 & - \\ \vspace{0.2cm}
Retail\_30 FIT\_6 & 0.05 & 0.25 & - & 0.06 & - \\ \vspace{0.2cm} 
Retail\_30 FIT\_4 & 0.05 & 0.25 & - & 0.04 & - \\ \vspace{0.2cm} 
Retail\_30 FIT\_2 & 0.05 & 0.25 & - & 0.02 & - \\ \vspace{0.2cm} 
Retail\_30 FIT\_0 & 0.05 & 0.25 & - & - &  No feed-in \\  
Retail\_30 FIT\_8 Cap & 0.05 & 0.25 & - & 0.08 &  $G^{pro2m}_{h}\leq \frac{1}{2}N_{pv}^{pro}$ \\ 
\midrule
\\
\multicolumn{6}{l}{\textbf{b) Scenarios with a fixed-part retail tariff}}  \\ [0.75ex]
Retail\_25 FIT\_8 & 0.05 & 0.20 & 250 & 0.08 & - \\ \vspace{0.2cm}
Retail\_20 FIT\_8 & 0.05 & 0.15 & 500 & 0.08 & - \\ \vspace{0.2cm}
Retail\_15 FIT\_8 & 0.05 & 0.10 & 750 & 0.08 & - \\ \vspace{0.2cm}
Retail\_25 FIT\_0 & 0.05 & 0.20 & 250 & - & No feed-in  \\ \vspace{0.2cm}
Retail\_20 FIT\_0 & 0.05 & 0.15 & 500 & - & No feed-in  \\ \vspace{0.2cm}
Retail\_15 FIT\_0 & 0.05 & 0.10 & 750 & - & No feed-in  \\ 
\midrule
\\
\multicolumn{6}{l}{\textbf{c) Scenarios with real-time pricing}}  \\ [0.75ex]
Retail\_30 FIT\_RTP & 0.05 & 0.25 & - & RTP &  - \\ \vspace{0.2cm}
Retail\_RTP FIT\_5 & RTP & 0.25 & - & 0.05 & - \\ \vspace{0.2cm}
Retail\_RTP FIT\_RTP & RTP & 0.25 & - & RTP & - \\ \vspace{0.2cm}
Retail\_RTP FIT\_RTP+3 & RTP & 0.25 & - & RTP+0.03 & - \\ 
\bottomrule 
\end{tabular}
\end{adjustbox}
\end{threeparttable}
\end{table}
\renewcommand{\baselinestretch}{1.4}

Scenario~$Retail\_30 \; FIT\_8$ is the baseline scenario. Retail tariff and FIT are comparable to the situation by 2019. The retail tariff for electricity consumption comprises a time-constant, volumetric energy charge~$t^{ener}_h$ and volumetric other charges~$t^{other}$. The energy charge of~$0.05$~EUR/kWh is calibrated endogenously according to the average energy wholesale market price~$\bar{\lambda}^{enbal}$ in an initial model calibration run without prosumage. The tariff component~$t^{other}$ of~$0.25$ EUR/kWh broadly reflects the level of non-energy charges contained in the German electricity retail price by 2019 (see appendix~\ref{app:sec:germany}). The volumetric consumption price for households sums up to~$0.30$~EUR/kWh. For a pure consumer household with an annual load of~$5$~MWh, this results in an annual expenditure of~$1500$~EUR on the electricity bill. The FIT is assumed to be~$0.08$~EUR/kWh, which is slightly above the projected LCOE of PV. 

The first group of scenarios keep the volumetric tariff price constant and vary the level of the feed-in tariff. Scenario~$Retail\_30 \; FIT\_0$ represents an extreme case in which grid feed-in of PV electricity by households is prohibited. 
Scenario $Retail\_30 \; FIT\_8 \; Cap$ is the same as the baseline, but additionally restricts the hourly maximum in-feed of prosumage PV energy into the grid to~$50$\% of the installed PV capacity. This is in line with the requirements for preferential loans in Germany by 2019. 

The second group of scenarios implement a greater fixed part for retail tariffs. Households pay an annual fixed fee~$t^{fix}$ and, in turn, a lower volumetric charge~$t^{other}$ for non-energy price components. This reflects a more capacity-oriented tariff design. For a pure consumer household with an annual load of~$5$~MWh, all scenarios with a fixed-part retail tariff result in the same annual bill of~$1500$~EUR as under the baseline and volumetric scenarios. 

The third group of scenarios represent dynamic pricing schemes. Households pay a volumetric retail tariff of~$0.30$~EUR/kWh in scenario $Retail\_30 \; FIT\_RTP$ and sell their electricity at the current wholesale market price, represented by the dual of the power sector energy balance~$\lambda^{enbal}_h$ in the model. Scenario~$Retail\_RTP \; FIT\_5$ assumes a time-varying energy price component of the retail tariff, in addition to a fixed volumetric component of~$0.25$~EUR/kWh. The feed-in tariff, in turn, is fixed at~$0.05$~EUR/kWh. The other two dynamic pricing scenarios impose a real-time price on both the retail and the feed-in sides, with an additional market premium of~$0.03$~EUR/kWh in scenario $Retail\_RTP \; FIT\_RTP+3$. This is motivated by the idea that the mean market value of PV energy is typically low in hours with high PV feed-in. The market premium may help to cover the cost difference between LCOE of PV and the wholesale market price.


\section{Results}\label{sec:results}
We show and interpret the model results with respect to household investments, how households use their prosumage systems, and several implications for the power sector. In doing so, we relate our findings to the prosumage incentives that households receive through price signals, as illustrated in figure~\ref{fig:ossenbrink}.

\subsection{Optimal household investments into PV and storage}\label{subsec:results_investment}
\begin{figure}[hbt!]
\centering 
\footnotesize
\includegraphics[height=9.5cm]{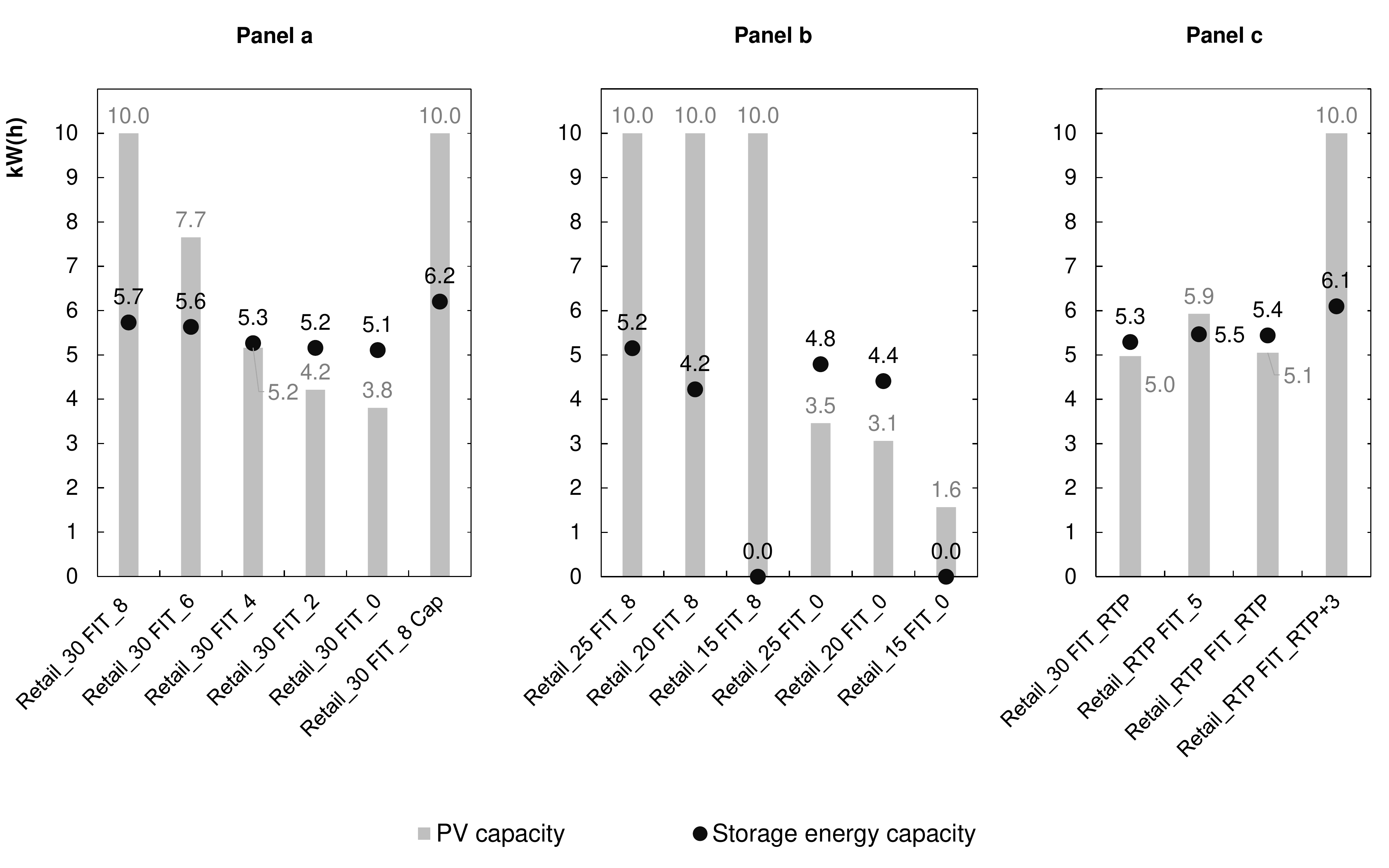}
\caption{Optimal PV and storage energy capacities of prosumage households.}\label{fig:optimal_capacities}
\end{figure}

In the baseline scenario~$Retail\_30 \; FIT\_8$, prosumage households install a PV capacity of~$10$~kW, storage energy capacity of 5.7~$kWh$, and storage power capacity of~$1.2$~kW (figure~\ref{fig:optimal_capacities}). The FIT is above the levelized cost of PV and grid feed-in never constitutes a net loss. The storage helps to substitute grid energy, priced at the volumetric retail tariff, by self-generated energy, at the levelized cost of PV and storage. Consequently, prosumage is profitable, and the baseline scenario refers to area~$F$ in figure~\ref{fig:ossenbrink}.

Panel~\ref{fig:optimal_capacities}a shows the results for the first group of scenarios. At a volumetric retail tariff of~$0.30$~EUR/kWh, a FIT below the LCOE of PV of~$0.08$~EUR/kWh yields lower optimal PV investments, referring to area~$E$ in figure~\ref{fig:ossenbrink}. Yet optimal battery energy capacities are relatively stable because the storage is built to optimize the amount of self-consumed PV generation given the retail price. Beyond a storage capacity of~$5$ to~$6$~kWh, the costs of additional storage become too high compared to the small increase in self-consumption and avoidance of grid consumption that can be reached. This finding also holds true in the absence of remuneration for grid feed-in. A feed-in cap yields somewhat higher optimal storage capacities to accommodate a greater share of self-consumption.

Results for the second group of scenarios follow a similar line of reasoning (panel~\ref{fig:optimal_capacities}b). A high feed-in tariff of~$0.08$~EUR/kWh, above the LCOE of PV, incentivizes maximum solar capacities. Without any grid feed-in, PV capacities are optimized only to serve self-consumption. Optimal battery capacities are largely governed by the design of the retail tariff and, accordingly, the profitability of self-consumption. Greater fixed parts, together with lower volumetric price components, trigger smaller optimal storage capacities. Eventually, the difference between the volumetric retail tariff and FIT is too small to profitably cover expenditure on the storage battery. Consequently, scenarios~$Retail\_15 \; FIT\_8$ and~$Retail\_15 \; FIT\_0$ refer to areas~$C$ and~$D$ in figure~\ref{fig:ossenbrink}, respectively.

If PV feed-in is remunerated by the real-time price, optimal PV capacities are below the maximum, at around~$5.5$~kW per household (panel~\ref{fig:optimal_capacities}c). In fact, the average price for which households sell electricity to the market is slightly above~$0.04$~EUR/kWh, rendering results comparable to scenario~$Retail\_30 \; FIT\_4$. A FIT of~$0.05$~EUR/kWh, or a market premium of~$0.03$~EUR/kWh, thus, yield greater optimal PV investments. Average real-time retail prices at which households buy electricity from the market are slightly below~$0.05$~EUR/kWh. With the volumetric component of~$0.25$~EUR/kWh, the eventual retail rate is around~$0.30$~EUR/kWh. Accordingly, optimal storage energy investments are comparable to the first group of scenarios and range between~$5$ and~$6$~kWh. 


\subsection{Optimal household dispatch: self-generation, self-consumption, and expenditure}\label{subsec:results_dispatch}
\begin{figure}[hbt!]
\centering 
\footnotesize
\includegraphics[height=8.5cm]{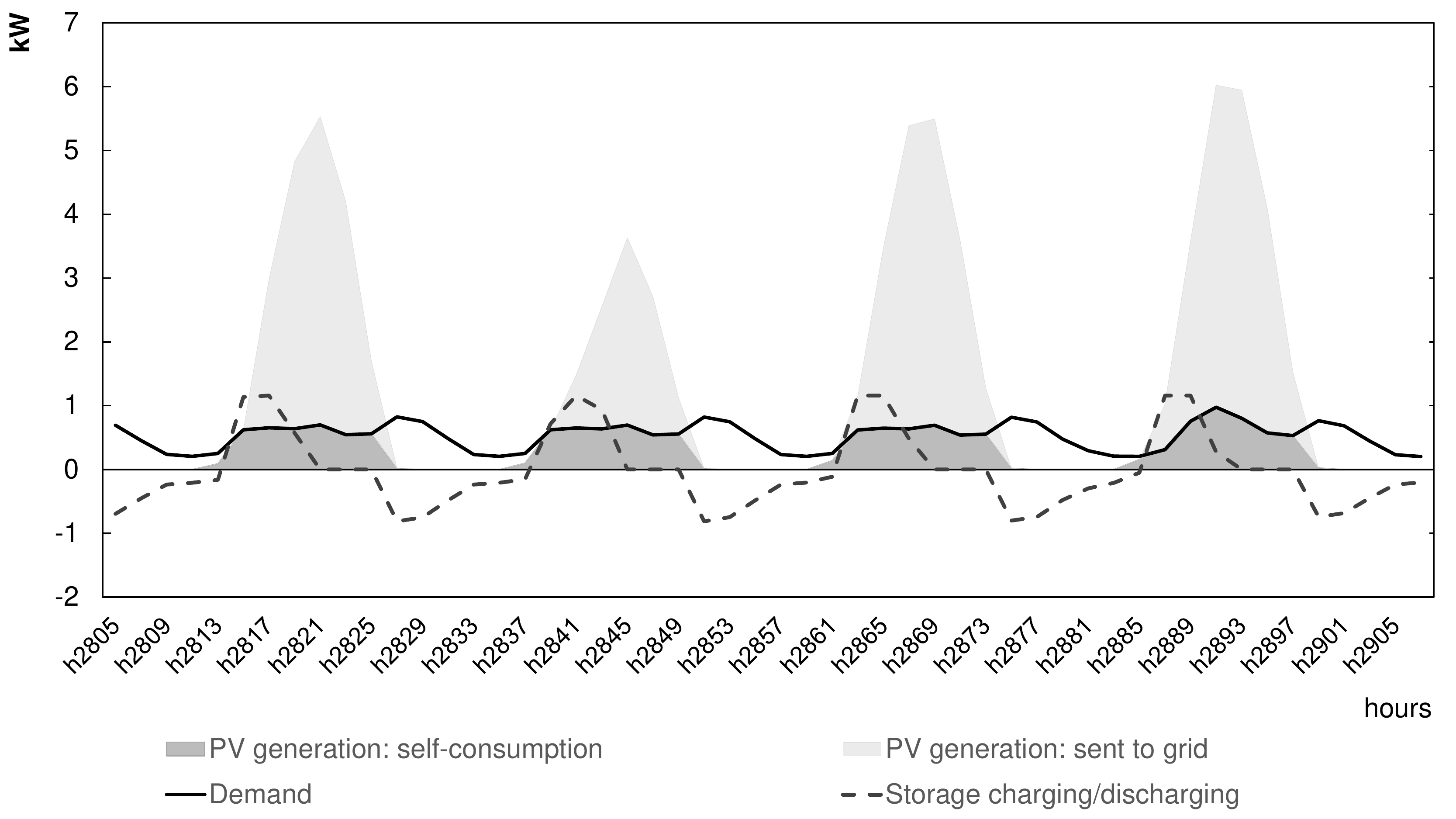}
\caption{Exemplary dispatch plot of prosumage household in baseline scenario.}\label{fig:pattern}
\end{figure}

We start with some intuition. Figure~\ref{fig:pattern} shows the dispatch behavior of a prosumage household for three sunny days at the end of April, taken from the baseline scenario. During the day, the available solar energy exceeds the demand of the prosumage household, and it can consume only a part of PV energy directly. Most of the PV energy is sent to the grid, peaking in hours of high solar radiation. Prosumage households charge their battery fully in the morning and discharge it in the evening when available PV energy declines. Both the feed-in and retail tariffs are time-invariant. Therefore, the household is not incentivized to schedule grid feed-in to hours with higher prices, for instance in the morning, and grid demand to hours with lower prices, for instance at night. 

\begin{figure}[hbt!]
\centering 
\footnotesize
\includegraphics[height=9.5cm]{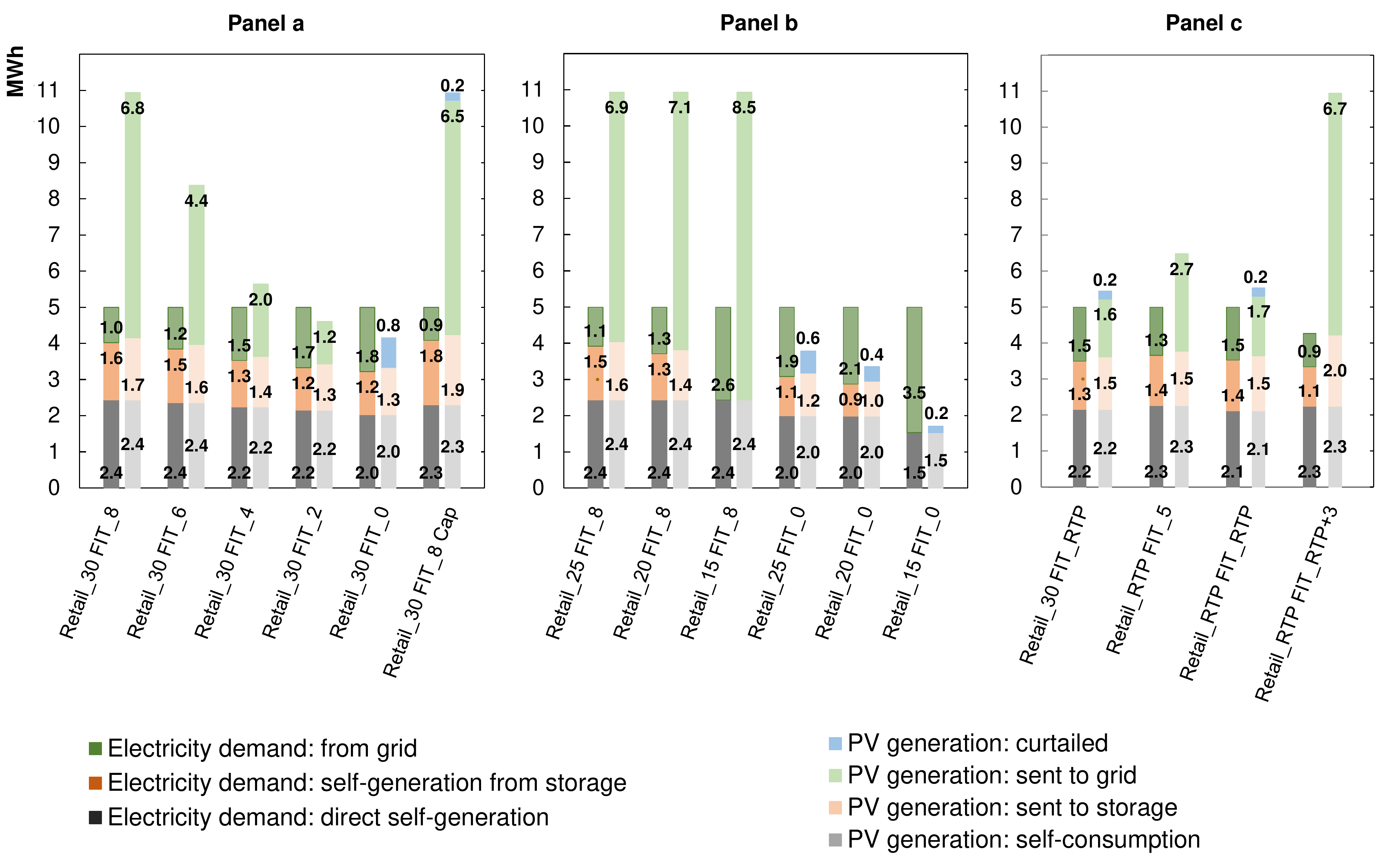}
\caption{Composition of prosumage households' electricity demand (left columns) and uses of households' PV electricity (right columns).}\label{fig:shares}
\end{figure}

Figure~\ref{fig:shares} shows how, depending on the tariff design, households use their PV electricity and from which sources they satisfy their electricity demand. In the baseline scenario, households satisfy~$4.0$~MWh of their annual electricity demand of~$5.0$~MWh with self-generated electricity (panel~\ref{fig:shares}a). This corresponds to an autarky rate of~$80$\%. Almost half of the annual demand ($2.4$~MWh) is directly satisfied with PV energy. Around~$30$\% ($1.6$~MWh) of demand is met with energy from the battery, while only one fifth is obtained from the grid (panel~\ref{fig:shares}a). The overall PV generation of~$10.9$~MWh exceeds annual demand, yielding a self-consumption rate slightly below~$40$\%. More than half of the PV generated electricity is sent to the grid. This does not change much if the baseline setting is combined with a cap on maximum PV feed-in.

For a lower FIT, the composition of sources that satisfy electricity demand stays relatively stable -- the autarky rate is between~$64$\% and~$78$\% -- with a somewhat increasing share of grid electricity (panel~\ref{fig:shares}a). Likewise, the \textit{absolute} volumes of self-consumption, direct or facilitated by the battery, only decrease slightly. Yet optimal PV panels are smaller and generate overall less electricity, yielding a higher \textit{relative} proportion of self-consumption. If grid feed-in is prohibited, households consume~$80$\% of their PV electricity themselves, with the remaining energy curtailed. 
\vspace{0.5cm}

For greater fixed parts in retail tariffs, and accordingly lower volumetric parts, panel~\ref{fig:shares}b shows a stronger dependency on the grid. In this group of scenarios, it is less attractive to substitute grid energy with self-generated energy. For illustration, compare the baseline $Retail\_30 \; FIT\_8$ with scenario~$Retail\_15 \; FIT\_8$. With a halved volumetric retail price, a prosumage household satisfies~$2.6$~MWh of her annual electricity demand from the grid, compared to only~$1.0$~MWh in the baseline. Accordingly, the autarky rate drops from~$80$\% to below~$50$\%. In the most extreme scenario~$Retail\_15 \; FIT\_0$, prosumage households satisfy only~$30$\% ($1.5$~MWh) of their demand on-site and source approximately~$70$\% ($3.5$~MWh) from the grid. Lower volumetric retail tariffs also decrease the volume of self-consumed energy. In the baseline, $4.1$~MWh out of~$10.9$~MWh PV generation are consumed on-site, that is, around~$38$\%. In scenario~$Retail\_15 \; FIT\_8$, only~$2.4$~MWh are consumed on-site, amounting to~about~$22$\%, and~$8.5$~MWh are fed into the grid. Battery-facilitated self-consumption does not take place.

Results for real-time pricing, with a mean price of around~$0.05$~EUR/kWh for energy consumption plus a volumetric non-energy charge of~$0.25$~EUR/kWh, are comparable to the fixed retail rate of~$0.30$~EUR/kWh (panel~\ref{fig:shares}c). Likewise, feed-in remuneration at real-time market prices of, on average, $0.04$~EUR/kWh results in a similar use pattern for PV electricity as under the scenario with a comparable fixed FIT. A market premium of~$0.03$~EUR/kWh increases the mean real-time feed-in price, with an accordingly greater grid feed-in. 

\begin{figure}[hbt!]
\centering 
\footnotesize
\includegraphics[height=9.5cm]{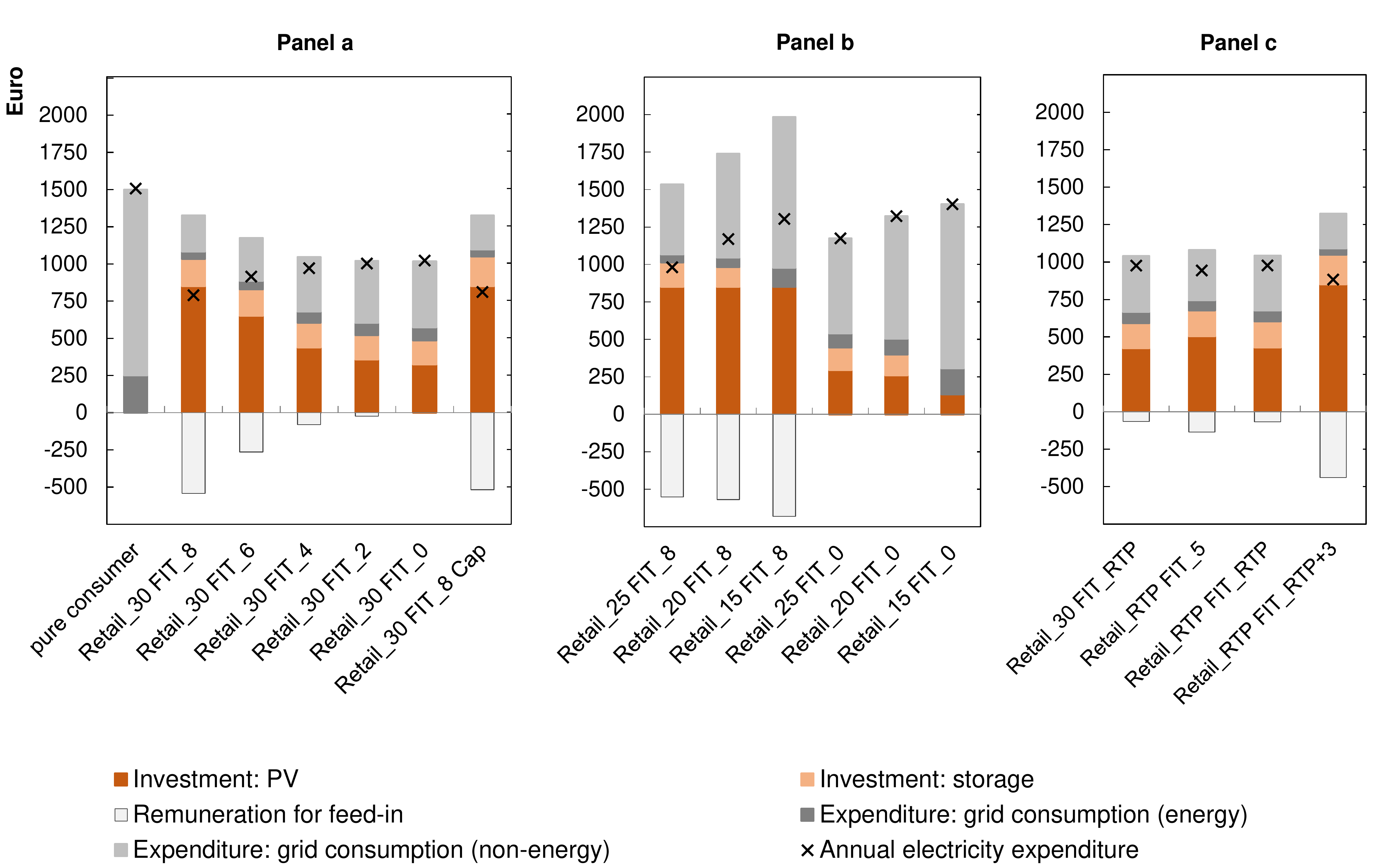}
\caption{Annual expenditures of a prosumage household on electricity, broken down into annualized investments into PV and storage capacities, costs for grid electricity, and revenues PV grid feed-in.}\label{fig:bill}
\end{figure}

Tariff design also affects the electricity bill of prosumage households (figure~\ref{fig:bill}). Prosumage households profit the most in scenarios where they offset large parts of their grid consumption. In the baseline~$Retail\_30 \; FIT\_8$, their annual bill on electricity expenditure is~$785$~EUR (panel~\ref{fig:bill}a). It includes annualized costs of the PV and storage systems, expenses for grid consumption, and revenues from PV power feed-in. Compared to a pure consumer, the bill almost halves. If the FIT is lower, total net expenditures rise slightly, with a pronounced shift toward expenses for grid consumption. If grid feed-in is capped at~$50$\% of the installed PV capacity, household expenditures are largely the same as under the baseline. This means that households can effectively adjust their energy feed-in without incurring curtailment losses.
\vspace{0.5cm}

Lower volumetric retail tariffs with higher fixed parts influence expenditures of prosumage households to a greater extent (panel~\ref{fig:bill}b). Since they must pay a fixed network charge in any case, the saving potential of prosumage compared to pure consumer behavior deteriorates. In any of the scenarios, non-energy payments for grid consumption constitute a substantial part of household expenditure. 

The real-time pricing scenarios (panel~\ref{fig:bill}c) only have moderate effects on the electricity bill when compared to the baseline. Prosumage households can still reduce their annual expenditure below~$1000$~EUR, representing a cut of more than a third over pure consumers. Differences between the real-time pricing scenarios are rather small because the assumed inelastic demand limits households' options to respond to price signals on the consumption side.


\subsection{Selected effects on the power sector}\label{subsec:results_system}
Beyond households, the prosumage tariffs implemented in the different scenarios also affect the power sector. We discuss implications for peak feed-in, PV generation, and recovery of non-energy costs. 


\subsubsection{Peak feed-in}\label{subsubsec:results_system_grid}
\begin{figure}[hbt!]
\centering
\footnotesize
\includegraphics[height=9.5cm]{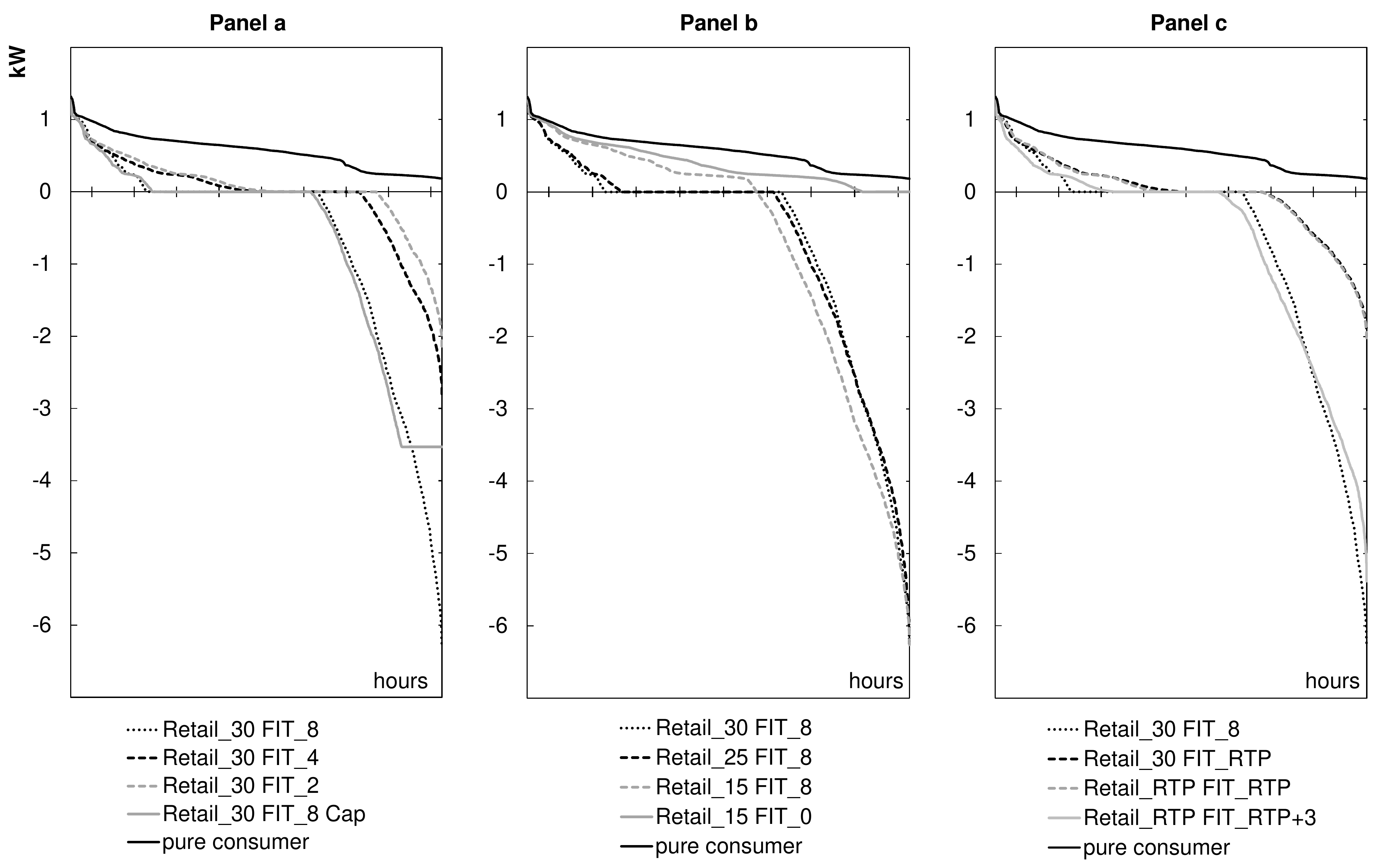}
\caption[Selected residual load duration curves of prosumage households]{Residual load duration curves of prosumage households for selected scenarios.}
\label{fig:results_res_load}
\end{figure}

While we do not model the electricity distribution grid explicitly and idiosyncratic configurations prevent general conclusions, peak demand and peak feed-in are suitable indicators: higher peaks incur higher stress. Figure~\ref{fig:results_res_load} shows residual load duration curves of a prosumage households for selected scenarios. A residual load duration curve is a graphical representation of residual load, that is, household net demand from the grid or net feed-in to the grid. It is ordered for all hours of a year in a descending fashion.   
\vspace{0.5cm}

In the baseline scenario~$Retail\_30 \; FIT\_8$, prosumage households consume electricity from the grid in~$20$\% of all hours (panel~\ref{fig:results_res_load}a). Their residual load is zero during~$46$\% of all hours, meaning that they satisfy their own electricity demand without sending surplus energy to the grid. In the remaining third of the year, prosumage households feed PV energy to the grid. For comparison, the upper dark gray lines indicate the residual load duration curve of a pure consumer. Her residual load is positive throughout the year because she consumes power from the grid at all times.

Both the sizes of the PV panel and battery shape the residual load duration curve. To this end, compare the baseline with scenario~$Retail\_15 \; FIT\_8$, in which households do not invest in storage but have the same PV capacities (panel~\ref{fig:results_res_load}b). In that case, there are no hours of zero residual load. Households consume grid energy~$60$\% of the time; the remaining~$40$\% of hours they feed surplus energy into the grid. The effect of PV capacity on the residual load duration curve can be inferred from comparing the baseline with scenario $Retail\_30 \; FIT\_2$, where prosumage households have a battery capacity of comparable size yet less than half of the PV capacity (panel~\ref{fig:results_res_load}a). This results in more hours of grid consumption, less hours of energy feed-in, and a lower absolute feed-in of surplus energy. 

The left-hand sides of all panels in figure~\ref{fig:results_res_load} indicate that none of the pricing schemes helps reducing peak residual demand. In all scenarios, it is around~$1.3$~kW. Thus, prosumage households do not shift their consumption patterns in a way that alleviates potential stress on the distribution grid. The right-hand sides of the residual load duration curves show the feed-in peaks. The peak is the higher the larger the PV capacity. In the baseline, it amounts to~$6.3$~kW; a similar order of magnitude emerges in the other scenarios with maximum PV investment. With~$3.5$~kW, it is about half this size if maximum feed-in is capped at~$50$\%. Note that this reduction is reached with only little financial disadvantage for prosumage households.

In the scenarios with real-time pricing on the generation side (panel~\ref{fig:results_res_load}c), the maximum feed-in is also somewhat lower, around~$2$~kW, compared to the baseline. Beyond the smaller PV size, this is also driven by low market prices in hours with high solar radiation. Thus, households have an incentive to avoid those hours for feeding into the grid. Comparing scenarios $Retail\_30 \; FIT\_4$ and $Retail\_30 \; FIT\_RTP$ illustrates this point. While PV and storage capacities are about the same size, the maximum feed-in is~$1$~kW higher for the time-invariant FIT. A similar rationale applies to the market premium scenario compared to the fixed FIT scenarios. Also here, maximum feed-in is lower by about~$1$~kW at similar capacities.


\subsubsection{Contribution to PV expansion and non-energy power system costs}\label{subsubsec:results_system_recovery}
\begin{figure}[hbt!]
\centering
\footnotesize
\includegraphics[height=9.5cm]{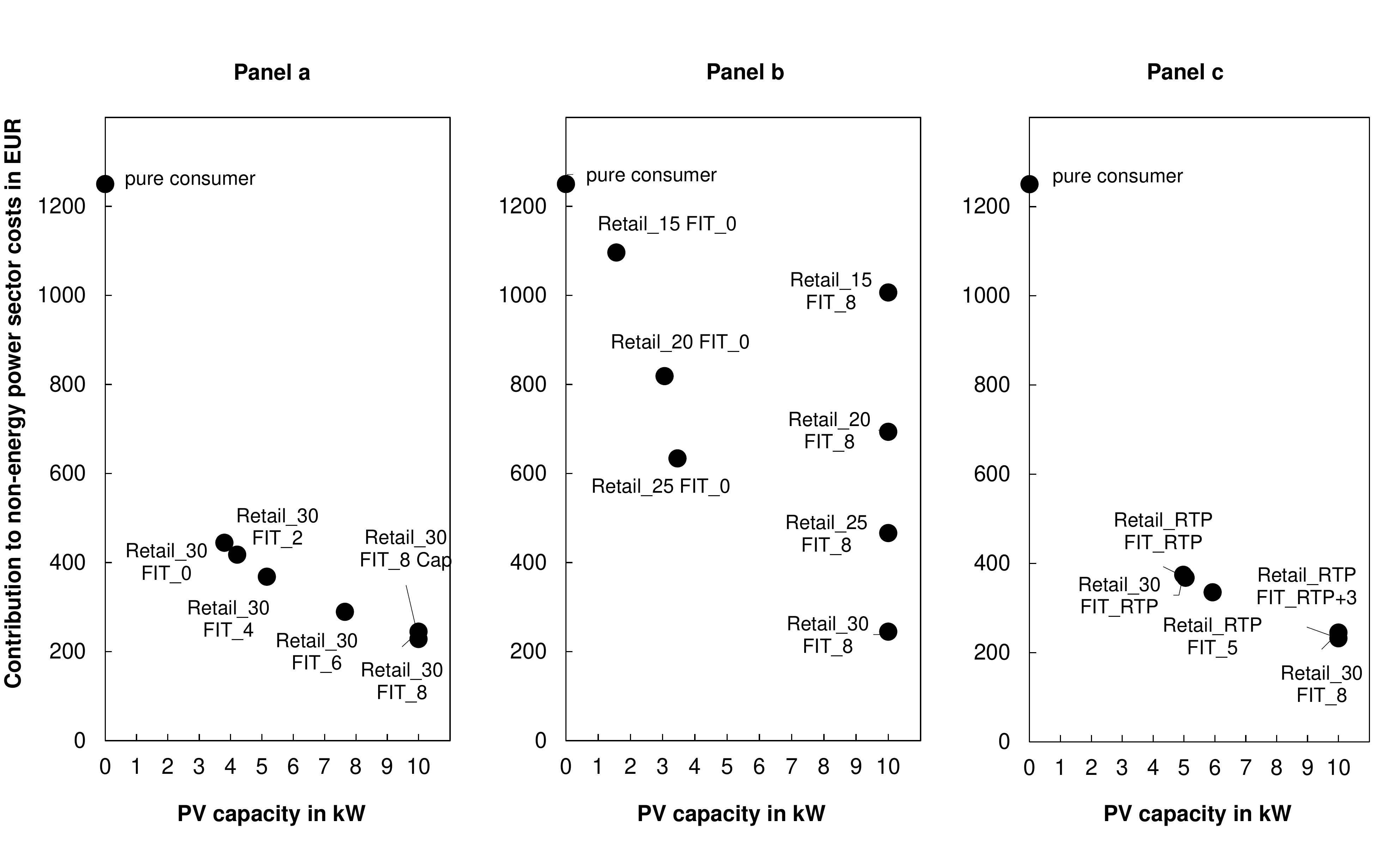}
\caption{Tradeoff between the annual contribution to non-energy power sector costs and PV capacities.}
\label{fig:results_tradeoff}
\end{figure}

Finally, we provide results how prosumage tariffs can impact the expansion of PV capacities on the one hand and recovery of non-energy costs of the power sector on the other. These non-energy system costs may account for grid costs, surcharges for financing renewables, and other fees. To this end, we quantify a prosumage household's contribution by summing up expenses on the volumetric and fixed charges $t^{other}$ and~$t^{fix}$. We contrast this figure with households' PV investments. Figure~\ref{fig:results_tradeoff} shows the power sector non-energy cost contribution on the horizontal axis. PV capacity, as contribution to providing renewable energy, are on the vertical axis. The further to the north-east, the better a scenario addresses both dimensions.

In the baseline scenario, prosumage households have an autarky rate of~$80$\% and contribute to non-energy power sector costs with about~$245$~EUR/year (panel~\ref{fig:results_tradeoff}a). For a pure consumer, this number amounts to~$1250$~EUR/year. Thus, the baseline tariff scheme may hamper cost recovery of energy infrastructures. It incentivizes households to lower grid consumption and save on the retail tariff with the according volumetric surcharges. Also in the other scenarios with dominant volumetric retail pricing, prosumage households contribute modestly to non-energy power system costs. In general, the higher the autarky rate and the less a household pays for volumetric price components, the less it contributes. Concerning the contribution to expanding renewable energy, the baseline features maximum PV investments of~$10$~kW. As discussed, lower FITs lead to lower PV investments and, thus, a lower contribution to expanding renewables. 

In contrast, if households face a fixed part in their retail tariff independent of their annual consumption level, as in the second group of scenarios, greater autarky does not necessarily go along with a lower contribution to non-energy power sector costs (panel~\ref{fig:results_tradeoff}b). The fixed payments allow households to save on volumetric retail expenses for energy, yet make sure that they contribute to non-energy system costs. As in all scenarios, a FIT above the LCOE of PV triggers maximum investments. Thus, these tariff designs appear most suitable to involve prosumage households in the recovery of fixed power system costs and incentivize large PV capacities. Results for the real-time pricing scenarios (panel~\ref{fig:results_tradeoff}c) do not differ much from the scenarios with a purely volumetric retail tariff.
\vspace{0.5cm}

\addtocounter{footnote}{+6} 


\section{Discussion of limitations}\label{sec:limitations}
Our analysis is subject to several limitations. First, the model does not endogenously capture all power sector costs related to prosumage. We can only give an indication whether a certain tariff setting may be detrimental to distribution (or transmission) grids; a deeper analysis would require explicitly modeling the underlying power flows. We also take on a static perspective. We set non-energy cost component of retail tariffs exogenously and abstract from their potential adaption over time. Hence, concerning utilities, the \textit{death spiral} effect is not explicitly accounted for. Concerning households, this hampers the analysis of rebound effects. Thus, determining an optimal tariff design is eventually not possible, and not aimed for, within the chosen modeling framework. 

Second, we apply several simplifications to the power sector dispatch problem. The model abstracts from inter-temporal dispatch restrictions like ramping constraints or minimum up and down times of thermal generators. This tends to overestimate the flexibility of conventional generators. Beyond pumped-hydro storage, we abstract from further flexibility options in the power sector level like flexible sector coupling. Together with our focus on Germany only, this tends to under-estimate flexibility and, accordingly, overestimate price volatility in the dynamic pricing scenarios.  

Third, we apply several simplifications to the household side. Model-wise, we abstract from the influence of uncertainty. However, we expect this to have little impact on results. It is plausible that necessary information on demand, solar radiation, and wholesale market prices is readily available from smart forecasting tools, and prosumage scheduling is close to optimal in shorter time frames. Data-wise, the standard load profile and national PV capacity factors are likely smoother than actual households' profiles. This tends to overestimate the match between household demand and PV generation, yielding higher autarky and self-consumption rates than observed in reality \citep{weniger_sizing_2014}. On a behavioral note, imperfect information or transaction costs may drive away household decisions from optimal values identified in our research. However, it is reasonable to assume households do, on average, choose a system according to their consumption needs and market incentives. When addressing these limitations, we expect our findings to be preserved, though potentially less pronounced than suggested by the model. Most importantly, we abstract from heterogeneous households. Therefore, we cannot explicitly derive distributional implications between different prosuming households as well as prosuming and non-prosuming households.

Last, cost assumptions for residential PV and storage batteries are important input parameters. If storage costs decline to a lesser extent than assumed or if interest rates increase, substituting grid energy with self-generated power will be less profitable. However, given the already high retail price in Germany, storage will still become profitable even if investment costs are considerably higher than in the model scenarios. As a consequence, optimal storage capacity would be smaller, but general findings would still apply since all scenario results would be affected in a fairly similar manner.\footnote{In a not reported sensitivity with double storage costs, prosumage households invest in a storage energy capacity of~$4$~kWh and a PV capacity of~$10$~kW. Only if we calibrate storage costs in line with the most conservative cost decrease prediction of~\citet{schmidt_future_2017}, self-consumption without batteries will be more profitable than prosumage in the model scenarios. Model results are less sensitive regarding the costs of PV systems.} 

Possible directions for future research could address these limitations. Specifically, a more detailed, endogenous representation of retail and network tariffs would enable a dynamic perspective on tariff formation and cross-subsidies between consumers and prosumage households. Moreover, future research could analyze the effects of different tariff designs when including demand-side management, heat pumps, and electric vehicles on the household side. It is likely that this additional flexibility and increased demand will intensify the observed differences between scenarios.

    
\section{Conclusions}\label{sec:conclusion}
Solar prosumage is still a niche phenomenon in most power markets as of 2019. However, as storage costs further decline and self-consumption of electricity becomes increasingly attractive, solar prosumage has the potential to unfold disruptive effects on the power sector. Whether it is attractive for households to invest in batteries and increase their level of PV self-generation vitally depends on the design of retail and feed-in tariffs.

Our model-based analysis for German 2030 scenarios shows that prosumage becomes profitable in many settings, even in absence of purchasing subsidies for batteries or a FIT for PV energy. Given established cost projections for PV and battery storage and a high volumetric retail price, households face strong incentives to substitute large parts of their grid energy consumption with self-generated energy. Departing from the current electricity tariff design, and introducing a greater fixed tariff part for households, would deteriorate incentives to invest in storage systems. 

At high feed-in tariffs, households invest in large PV capacities, reach considerable shares of self-generation, and feed a good amount of surplus PV energy into the grid. In contrast, households opt for smaller PV capacities at low feed-in remunerations, since they then face a trade-off regarding the optimal PV system size. Yet the optimal storage capacity for a prosumage household is less sensitive to the feed-in tariff schemes. This is because the storage capacity is driven by the time profiles of residual residential electricity demand and PV generation: beyond a certain storage capacity, the costs of increasing self-generation through additional storage become prohibitive.

From an energy (transition) policy perspective, prosumage is not desirable or detrimental as such. Potentially unintended consequences can arise if high volumetric retail and FIT pricing schemes prevail. High volumetric retail tariffs give prosumage households the incentive to reach high levels of autarky. While these households save costs through self-generation, they contribute less to the national total of grid expenses, renewable surcharges or taxes. Especially concerning grid costs, such revenue shortfalls eventually must be covered by other, non-privileged electricity consumers, giving rise to distributional issues. This is aggravated by the conjuncture that prosumage households could induce distribution grid overloads because of peak PV feed-in and thus over-proportionately contribute to network instability and costs. 

While a high feed-in tariff increases the renewable energy capacity provided by the residential sector, it does not convey incentives for energy-market-oriented dispatch of PV-plus-battery systems. Real-time pricing could incentivize prosumage households to better align their self-consumption patterns with the electricity wholesale market. A maximum feed-in policy, which caps the peak energy fed into the grid, seems to be suitable to achieve distribution grid reliefs, without causing a large financial disadvantage for prosumage households.

Taken together, none of the tariff design options investigated here seem to dominate in the investigated respects. A greater contribution to non-energy power sector costs can generally be achieved with a greater role for fixed non-energy charges. This can be combined with a FIT, and potentially also with a feed-in cap. Given the FIT is high enough, it incentivizes high investments into PV capacity. At the same time, it is relatively easy to implement from an administrative and ICT perspective. Such a combination of FITs and higher fixed retail tariff parts may thus help to foster the transition to renewable energies and, at the same time, to keep up households' contribution to recover non-energy power sector costs.


\newpage
\pagebreak
\bibliographystyle{plainnat}
\bibliography{references}

\begin{thebibliography}{63}
\providecommand{\natexlab}[1]{#1}
\providecommand{\url}[1]{\texttt{#1}}
\expandafter\ifx\csname urlstyle\endcsname\relax
  \providecommand{\doi}[1]{doi: #1}\else
  \providecommand{\doi}{doi: \begingroup \urlstyle{rm}\Url}\fi

\bibitem[Agnew and Dargusch(2015)]{agnew_effect_2015}
Scott Agnew and Paul Dargusch.
\newblock Effect of residential solar and storage on centralized electricity
  supply systems.
\newblock \emph{Nature Climate Change}, 5\penalty0 (4):\penalty0 315, 2015.
\newblock \doi{10.1038/nclimate2523}.

\bibitem[Bazilian et~al.(2013)Bazilian, Onyeji, Liebreich, MacGill, Chase,
  Shah, Gielen, Arent, Landfear, and Zhengrong]{bazilian_2013}
Morgan Bazilian, Ijeoma Onyeji, Michael Liebreich, Ian MacGill, Jennifer Chase,
  Jigar Shah, Dolf Gielen, Doug Arent, Doug Landfear, and Shi Zhengrong.
\newblock Re-considering the economics of photovoltaic power.
\newblock \emph{Renewable Energy}, 53:\penalty0 329 -- 338, 2013.
\newblock \doi{10.1016/j.renene.2012.11.029}.

\bibitem[BDEW(2015)]{bdew_standardlastprofile_2015}
BDEW.
\newblock Standardlastprofile {Strom}. {Standardlastprofil} {Haushalt} 2015,
  2015.
\newblock URL \url{https://www.stromnetz.berlin/netz-nutzen/netznutzer}.
\newblock Access date: 04.12.2018.

\bibitem[Bertsch et~al.(2017)Bertsch, Geldermann, and Lühn]{bertsch_what_2017}
Valentin Bertsch, Jutta Geldermann, and Tobias Lühn.
\newblock What drives the profitability of household {PV} investments,
  self-consumption and self-sufficiency?
\newblock \emph{Applied Energy}, 204:\penalty0 1--15, 2017.
\newblock \doi{https://doi.org/10.1016/j.apenergy.2017.06.055}.

\bibitem[BNetzA(2018{\natexlab{a}})]{bnetza_eeg-registerdaten_2018}
BNetzA.
\newblock {EEG}-{Re}­gis­ter­da­ten und -{För}­der­sät­ze.
  {Fördersätze} für {PV}-{Anlagen}. {Bundesnetzagentur (BNetzA)},
  2018{\natexlab{a}}.
\newblock URL
  \url{https://www.bundesnetzagentur.de/DE/Sachgebiete/ElektrizitaetundGas/Unternehmen_Institutionen/ErneuerbareEnergien/ZahlenDatenInformationen/EEG_Registerdaten/EEG_Registerdaten_node.html}.
\newblock Access date: 01.12.2018.

\bibitem[BNetzA(2018{\natexlab{b}})]{bnetza_genehmigung_2018}
BNetzA.
\newblock Genehmigung des {Szenariorahmens} 2019-2030. {Juni} 2018.
  {Bundesnetzagentur (BNetzA)}, 2018{\natexlab{b}}.
\newblock URL
  \url{https://www.netzausbau.de/bedarfsermittlung/2030_2019/szenariorahmen2019-2030/de.html}.
\newblock Access date: 01.12.2018.

\bibitem[{BSW Solar}(2018{\natexlab{a}})]{bsw_solar_meilenstein_2018}
{BSW Solar}.
\newblock Meilenstein der {Energiewende}: 100.000ster {Solarstromspeicher}
  installiert. {Press release} 28.08.2018, 2018{\natexlab{a}}.
\newblock URL
  \url{https://www.solarwirtschaft.de/presse/pressemeldungen/pressemeldungen-im-detail/news/meilenstein-der-energiewende-100000ster-solarstromspeicher-installiert.html}.
\newblock Access date: 14.12.2018.

\bibitem[{BSW Solar}(2018{\natexlab{b}})]{bsw_solar_statistische_2018}
{BSW Solar}.
\newblock Statistische {Zahlen} der deutschen {Solarstrombranche}
  ({Photovoltaik}). {February} 2018, 2018{\natexlab{b}}.
\newblock URL
  \url{https://www.solarwirtschaft.de/fileadmin/user_upload/bsw_faktenblatt_pv_4018_4.pdf}.
\newblock Access date: 07.12.2018.

\bibitem[{Bundesministerium für Wirtschaft und Energie
  (BMWi)}(2018)]{eeg_in_zahlen_2018}
{Bundesministerium für Wirtschaft und Energie (BMWi)}.
\newblock {EEG in Zahlen: Vergütungen, Differenzkosten und EEG-Umlage 2000 bis
  2019}., 2018.
\newblock URL
  \url{https://www.erneuerbare-energien.de/EE/Navigation/DE/Recht-Politik/Das_EEG/DatenFakten/daten-und-fakten.html}.
\newblock Access date: 04.06.2019.

\bibitem[Costello and Hemphill(2014)]{costello_2014}
Kenneth~W. Costello and Ross~C. Hemphill.
\newblock Electric utilities’ ‘death spiral’: {Hyperbole} or reality?
\newblock \emph{The Electricity Journal}, 27\penalty0 (10):\penalty0 7--26,
  2014.
\newblock \doi{10.1016/j.tej.2014.09.011}.

\bibitem[Cucchiella et~al.(2016)Cucchiella, D'Adamo, and
  Gastaldi]{cucchiella_photovoltaic_2016}
Federica Cucchiella, Idiano D'Adamo, and Massimo Gastaldi.
\newblock Photovoltaic energy systems with battery storage for residential
  areas: {An} economic analysis.
\newblock \emph{Journal of Cleaner Production}, 131:\penalty0 460--474, 2016.
\newblock \doi{10.1016/j.jclepro.2016.04.157}.

\bibitem[Darghouth et~al.(2016)Darghouth, Wiser, Barbose, and
  Mills]{darghouth_net_2016}
Naïm~R. Darghouth, Ryan~H. Wiser, Galen Barbose, and Andrew~D. Mills.
\newblock Net metering and market feedback loops: {Exploring} the impact of
  retail rate design on distributed {PV} deployment.
\newblock \emph{Applied Energy}, 162:\penalty0 713--722, 2016.
\newblock \doi{10.1016/j.apenergy.2015.10.120}.

\bibitem[(Destatis)(2018)]{statistisches_bundesamt_destatis_stromabsatz_2018}
Statistisches~Bundesamt (Destatis).
\newblock Stromabsatz und {Erlöse} der {Elektrizitätsversorgungs}-
  unternehmen: {Deutschland}, {Jahre}, {Abnehmergruppen}, 2018.
\newblock URL \url{www.destatis.de}.
\newblock Access date: 02.11.2018.

\bibitem[Dietrich and Weber(2018)]{dietrich_what_2018}
Andreas Dietrich and Christoph Weber.
\newblock What drives profitability of grid-connected residential {PV} storage
  systems? {A} closer look with focus on {Germany}.
\newblock \emph{Energy Economics}, 74:\penalty0 399--416, 2018.
\newblock \doi{10.1016/j.eneco.2018.06.014}.

\bibitem[DIHK(2018)]{dihk_merkblatt_2018}
DIHK.
\newblock Merkblatt kleine {PV}-{Anlagen}: {Hinweise} zum {Betrieb} einer
  {PV}-{Anlage} und zum {Ende} der {Förderdauer} nach dem {EEG}, 2018.
\newblock URL
  \url{https://www.dihk.de/themenfelder/innovation-und-umwelt/energie/energiewende/service/merkblatt-kleine-pv}.
\newblock Access date: 13.12.2018.

\bibitem[Dirkse and Ferris(1995)]{dirkse_path_1995}
Steven~P. Dirkse and Michael~C. Ferris.
\newblock The path solver: a nommonotone stabilization scheme for mixed
  complementarity problems.
\newblock \emph{Optimization Methods and Software}, 5\penalty0 (2):\penalty0
  123--156, 1995.
\newblock \doi{10.1080/10556789508805606}.

\bibitem[Eid et~al.(2014)Eid, Guillén, Marín, and
  Hakvoort]{eid_economic_2014}
Cherrelle Eid, Javier~Reneses Guillén, Pablo~Frías Marín, and Rudi Hakvoort.
\newblock The economic effect of electricity net-metering with solar {PV}:
  {Consequences} for network cost recovery, cross subsidies and policy
  objectives.
\newblock \emph{Energy Policy}, 75:\penalty0 244--254, 2014.
\newblock \doi{10.1016/j.enpol.2014.09.011}.

\bibitem[ENTSOE(2018)]{entsoe_tyndp_2018}
ENTSOE.
\newblock {TYNDP} 2018. {Scenario} {Report}. {Main} report, 2018.
\newblock URL
  \url{https://docstore.entsoe.eu/Documents/TYNDP%20documents/TYNDP2018/Scenario_Report_2018_Final.pdf.}
\newblock Access date: 14.01.2019.

\bibitem[{European Commission Joint Research Center, Institute for Energy and
  Transport (JRC)}(2014)]{etri_2014}
{European Commission Joint Research Center, Institute for Energy and Transport
  (JRC)}.
\newblock {ETRI 2014 (Energy Technology Reference Indicator projections for
  2010 - 2050)}, 2014.
\newblock URL \url{https://setis.ec.europa.eu/system/files/ETRI_2014.pdf}.
\newblock Access date: 04.06.2019.

\bibitem[Facchinei and Pang(2007)]{facchinei_finite-dimensional_2007}
Francisco Facchinei and Jong-Shi Pang.
\newblock \emph{Finite-dimensional variational inequalities and complementarity
  problems}.
\newblock Springer Science \& Business Media, 2007.

\bibitem[Figgener et~al.(2018)Figgener, Haberschusz, Kairies, Wessels, Tepe,
  Ebbert, Herzog, and Sauer]{figgener_wissenschaftliches_2018}
Jan Figgener, David Haberschusz, Kai-Philipp Kairies, Oliver Wessels, Benedikt
  Tepe, Markus Ebbert, Reiner Herzog, and Dirk~Uwe Sauer.
\newblock Wissenschaftliches {Mess}- und {Evaluierungsprogramm}
  {Solarstromspeicher} 2.0, {Jahresbericht} 2018, 2018.
\newblock URL
  \url{http://www.speichermonitoring.de/fileadmin/user_upload/Speichermonitoring_Jahresbericht_2018_ISEA_RWTH_Aachen.pdf}.
\newblock Access date: 16.05.2019.

\bibitem[{Fraunhofer ISE}(2019)]{fraunhofer_ise_photovoltaics_2019}
{Fraunhofer ISE}.
\newblock Photovoltaics {Report}. {March} 2019, 2019.
\newblock URL
  \url{https://www.ise.fraunhofer.de/content/dam/ise/de/documents/publications/studies/Photovoltaics-Report.pdf}.
\newblock Access date: 04.06.2019.

\bibitem[Gautier et~al.(2019)Gautier, Hoet, Jacqmin, and {Van
  Driessche}]{gautier_2019}
Axel Gautier, Brieuc Hoet, Julien Jacqmin, and Sarah {Van Driessche}.
\newblock Self-consumption choice of residential {PV} owners under
  net-metering.
\newblock \emph{Energy Policy}, 128:\penalty0 648--653, 2019.
\newblock \doi{10.1016/j.enpol.2019.01.055}.

\bibitem[Green and Staffell(2017)]{green_prosumage_2017}
Richard Green and Iain Staffell.
\newblock “{Prosumage}” and the {British} electricity market.
\newblock \emph{Economics of Energy \& Environmental Policy}, 6\penalty0
  (1):\penalty0 33--49, 2017.
\newblock \doi{10.5547/2160-5890.6.1.rgre}.

\bibitem[Hinz et~al.(2018)Hinz, Schmidt, and Möst]{hinz_regional_2018}
Fabian Hinz, Matthew Schmidt, and Dominik Möst.
\newblock Regional distribution effects of different electricity network tariff
  designs with a distributed generation structure: {The} case of {Germany}.
\newblock \emph{Energy Policy}, 113:\penalty0 97--111, 2018.
\newblock \doi{10.1016/j.enpol.2017.10.055}.

\bibitem[Hoppmann et~al.(2014)Hoppmann, Volland, Schmidt, and
  Hoffmann]{hoppmann_economic_2014}
Joern Hoppmann, Jonas Volland, Tobias~S. Schmidt, and Volker~H. Hoffmann.
\newblock The economic viability of battery storage for residential solar
  photovoltaic systems–{A} review and a simulation model.
\newblock \emph{Renewable and Sustainable Energy Reviews}, 39:\penalty0
  1101--1118, 2014.
\newblock \doi{10.1016/j.rser.2014.07.068}.

\bibitem[Hughes and Bell(2006)]{hughes_compensating_2006}
Larry Hughes and Jeff Bell.
\newblock Compensating customer-generators: a taxonomy describing methods of
  compensating customer-generators for electricity supplied to the grid.
\newblock \emph{Energy Policy}, 34\penalty0 (13):\penalty0 1532--1539, 2006.
\newblock \doi{10.1016/j.enpol.2004.11.002}.

\bibitem[IEA(2018)]{iea_market_2018}
IEA.
\newblock Market {Report} {Series}: {Renewables} 2018. {Analysis} and
  {Forecasts} to 2023. {International} {Energy} {Agency}, 2018.
\newblock URL \url{https://www.iea.org/renewables2018/power/}.
\newblock Access date: 08.07.2019.

\bibitem[Kaschub et~al.(2016)Kaschub, Jochem, and Fichtner]{kaschub_solar_2016}
Thomas Kaschub, Patrick Jochem, and Wolf Fichtner.
\newblock Solar energy storage in {German} households: profitability, load
  changes and flexibility.
\newblock \emph{Energy Policy}, 98:\penalty0 520--532, 2016.
\newblock \doi{10.1016/j.enpol.2016.09.017}.

\bibitem[Khalilpour and Vassallo(2015)]{khalilpour_leaving_2015}
Rajab Khalilpour and Anthony Vassallo.
\newblock Leaving the grid: {An} ambition or a real choice?
\newblock \emph{Energy Policy}, 82:\penalty0 207--221, 2015.
\newblock \doi{10.1016/j.enpol.2015.03.005}.

\bibitem[Kubli(2018)]{kubli_squaring_2018}
Merla Kubli.
\newblock Squaring the sunny circle? {On} balancing distributive justice of
  power grid costs and incentives for solar prosumers.
\newblock \emph{Energy Policy}, 114:\penalty0 173--188, 2018.
\newblock \doi{10.1016/j.enpol.2017.11.054}.

\bibitem[Laws et~al.(2017)Laws, Epps, Peterson, Laser, and
  Wanjiru]{laws_utility_2017}
Nicholas~D. Laws, Brenden~P. Epps, Steven~O. Peterson, Mark~S. Laser, and
  G.~Kamau Wanjiru.
\newblock On the utility death spiral and the impact of utility rate structures
  on the adoption of residential solar photovoltaics and energy storage.
\newblock \emph{Applied Energy}, 185:\penalty0 627--641, 2017.
\newblock \doi{10.1016/j.apenergy.2016.10.123}.

\bibitem[Luthander et~al.(2015)Luthander, Widén, Nilsson, and
  Palm]{luthander_photovoltaic_2015}
Rasmus Luthander, Joakim Widén, Daniel Nilsson, and Jenny Palm.
\newblock Photovoltaic self-consumption in buildings: {A} review.
\newblock \emph{Applied Energy}, 142:\penalty0 80--94, 2015.
\newblock \doi{10.1016/j.apenergy.2014.12.028}.

\bibitem[Marwitz and Elsland(2018)]{marwitz_techno-economic_2018}
Simon Marwitz and Rainer Elsland.
\newblock Techno-economic modelling of low-voltage networks: {A} concept to
  determine the grid investment required in {Germany} and the implications for
  grid utilisation fees.
\newblock Technical report, Fraunhofer ISI, Working Paper Sustainability and
  Innovation, 2018.
\newblock URL
  \url{http://publica.fraunhofer.de/eprints/urn_nbn_de_0011-n-5153021.pdf}.
\newblock Access date: 08.07.2019.

\bibitem[Moshövel et~al.(2015)Moshövel, Kairies, Magnor, Leuthold, Bost,
  Gährs, Szczechowicz, Cramer, and Sauer]{moshovel_analysis_2015}
Janina Moshövel, Kai-Philipp Kairies, Dirk Magnor, Matthias Leuthold, Mark
  Bost, Swantje Gährs, Eva Szczechowicz, Moritz Cramer, and Dirk~Uwe Sauer.
\newblock Analysis of the maximal possible grid relief from {PV}-peak-power
  impacts by using storage systems for increased self-consumption.
\newblock \emph{Applied Energy}, 137:\penalty0 567--575, 2015.
\newblock \doi{10.1016/j.apenergy.2014.07.021}.

\bibitem[Muenzel et~al.(2015)Muenzel, Mareels, de~Hoog, Vishwanath,
  Kalyanaraman, and Gort]{muenzel_pv_2015}
Valentin Muenzel, Iven Mareels, Julian de~Hoog, Arun Vishwanath, Shivkumar
  Kalyanaraman, and Andrew Gort.
\newblock {PV} generation and demand mismatch: {Evaluating} the potential of
  residential storage.
\newblock In \emph{Innovative {Smart} {Grid} {Technologies} {Conference}
  ({ISGT}), 2015 {IEEE} {Power} \& {Energy} {Society}}, pages 1--5. IEEE, 2015.
\newblock \doi{10.1109/ISGT.2015.7131849}.

\bibitem[Neetzow et~al.(2019)Neetzow, Mendelevitch, and Siddiqui]{neetzow_2019}
Paul Neetzow, Roman Mendelevitch, and Sauleh Siddiqui.
\newblock Modeling coordination between renewables and grid: Policies to
  mitigate distribution grid constraints using residential pv-battery systems.
\newblock \emph{Energy Policy}, 132:\penalty0 1017--1033, 2019.
\newblock \doi{10.1016/j.enpol.2019.06.024}.

\bibitem[OPSD(2018)]{opsd_open_2018}
OPSD.
\newblock Open {Power} {System} {Data}. 2018. {Data} {Package} {Renewable}
  power plants. {Version} 2018-03-08, 2018.
\newblock URL
  \url{https://data.open-power-system-data.org/renewable_power_plants/2018-03-08}.
\newblock Access date: 25.11.2018.

\bibitem[Ossenbrink(2017)]{ossenbrink_how_2017}
Jan Ossenbrink.
\newblock How feed-in remuneration design shapes residential {PV} prosumer
  paradigms.
\newblock \emph{Energy Policy}, 108:\penalty0 239--255, 2017.
\newblock \doi{10.1016/j.enpol.2017.05.030}.

\bibitem[Pape et~al.(2014)Pape, Gerhardt, Härtel, Scholz, Schwinn, Drees,
  Maaz, Sprey, Breuer, and Moser]{pape_roadmap_2014}
Carsten Pape, Norman Gerhardt, Philipp Härtel, Angela Scholz, Rainer Schwinn,
  Tim Drees, Andreas Maaz, Jens Sprey, Christopher Breuer, and Albert Moser.
\newblock Roadmap {Speicher}. {Endbericht}. {Fraunhofer} {IWES}, {IAEW},
  {Stiftung} {Umweltenergierecht}. {November} 2014.
\newblock \emph{Fraunhofer IWES, Kassel}, 2014.
\newblock URL \url{http://publica.fraunhofer.de/documents/N-316127.html}.
\newblock Access date: 16.01.2019.

\bibitem[Pfenninger and Staffell(2016)]{pfenninger_long-term_2016}
Stefan Pfenninger and Iain Staffell.
\newblock Long-term patterns of {European} {PV} output using 30 years of
  validated hourly reanalysis and satellite data.
\newblock \emph{Energy}, 114:\penalty0 1251--1265, 2016.
\newblock \doi{10.1016/j.energy.2016.08.060}.

\bibitem[Picciariello et~al.(2015)Picciariello, Vergara, Reneses, Frías, and
  Söder]{picciariello_electricity_2015}
Angela Picciariello, Claudio Vergara, Javier Reneses, Pablo Frías, and Lennart
  Söder.
\newblock Electricity distribution tariffs and distributed generation:
  {Quantifying} cross-subsidies from consumers to prosumers.
\newblock \emph{Utilities Policy}, 37:\penalty0 23--33, 2015.
\newblock \doi{10.1016/j.jup.2015.09.007}.

\bibitem[Prognos(2016)]{prognos_eigenversorgung_2016}
Prognos.
\newblock Eigenversorgung aus {Solaranlagen}. {Das} {Potenzial} für
  {Photovoltaik}-{Speicher}-{Systeme} in {Ein}- und {Zweifamilienhäusern},
  {Landwirtschaft} sowie im {Lebensmittelhandel}. {Analyse} im {Auftrag} von
  {Agora} {Energiewende}., 2016.
\newblock URL
  \url{https://www.agora-energiewende.de/fileadmin2/Projekte/2016/Dezentralitaet/Agora_Eigenversorgung_PV_web-02.pdf}.
\newblock Access date: 05.12.2018.

\bibitem[Prol and Steininger(2017)]{prol_photovoltaic_2017}
Javier~López Prol and Karl~W. Steininger.
\newblock Photovoltaic self-consumption regulation in {Spain}: {Profitability}
  analysis and alternative regulation schemes.
\newblock \emph{Energy Policy}, 108:\penalty0 742--754, 2017.
\newblock \doi{10.1016/j.enpol.2017.06.019}.

\bibitem[Roulot and Raineri(2018)]{roulot_2018}
Jonathan Roulot and Ricardo Raineri.
\newblock The impacts of photovoltaic electricity self-consumption on value
  transfers between private and public stakeholders in {France}.
\newblock \emph{Energy Policy}, 122:\penalty0 459--473, 2018.
\newblock \doi{10.1016/j.enpol.2018.07.035}.

\bibitem[Say et~al.(2018)Say, John, Dargaville, and Wills]{say_coming_2018}
Kelvin Say, Michele John, Roger Dargaville, and Raymond~T. Wills.
\newblock The coming disruption: {The} movement towards the customer renewable
  energy transition.
\newblock \emph{Energy Policy}, 123:\penalty0 737--748, 2018.
\newblock \doi{10.1016/j.enpol.2018.09.026}.

\bibitem[Schill et~al.(2017)Schill, Zerrahn, and Kunz]{schill_prosumage_2017}
Wolf-Peter Schill, Alexander Zerrahn, and Friedrich Kunz.
\newblock Prosumage of solar electricity: pros, cons, and the system
  perspective.
\newblock \emph{Economics of Energy \& Environmental Policy}, 6\penalty0 (1),
  2017.
\newblock \doi{10.5547/2160-5890.6.1.wsch}.

\bibitem[Schill et~al.(2019)Schill, Zerrahn, and Kunz]{schill_2019}
Wolf-Peter Schill, Alexander Zerrahn, and Friedrich Kunz.
\newblock Solar prosumage: An economic discussion of challenges and
  opportunities.
\newblock In Jens Lowitzsch, editor, \emph{Energy Transition: Financing
  Consumer Co-Ownership in Renewables}, pages 703--731. Springer International
  Publishing, 2019.
\newblock \doi{10.1007/978-3-319-93518-8_29}.

\bibitem[Schittekatte et~al.(2018)Schittekatte, Momber, and
  Meeus]{schittekatte_future-proof_2018}
Tim Schittekatte, Ilan Momber, and Leonardo Meeus.
\newblock Future-proof tariff design: recovering sunk grid costs in a world
  where consumers are pushing back.
\newblock \emph{Energy Economics}, 70:\penalty0 484--498, 2018.
\newblock \doi{10.1016/j.eneco.2018.01.028}.

\bibitem[Schmidt et~al.(2017)Schmidt, Hawkes, Gambhir, and
  Staffell]{schmidt_future_2017}
Oliver Schmidt, Adam Hawkes, Ajay Gambhir, and Iain Staffell.
\newblock The future cost of electrical energy storage based on experience
  rates.
\newblock \emph{Nature Energy}, 2:\penalty0 17110, July 2017.
\newblock \doi{10.1038/nenergy.2017.110}.

\bibitem[Schröder et~al.(2013)Schröder, Kunz, Meiss, Mendelevitch, and
  Von~Hirschhausen]{schroder_current_2013}
Andreas Schröder, Friedrich Kunz, Jan Meiss, Roman Mendelevitch, and Christian
  Von~Hirschhausen.
\newblock Current and prospective costs of electricity generation until 2050.
\newblock Technical report, DIW Data Documentation, 2013.
\newblock URL
  \url{https://www.diw.de/documents/publikationen/73/diw_01.c.424566.de/diw_datadoc_2013-068.pdf}.

\bibitem[Simshauser(2016)]{simshauser_distribution_2016}
Paul Simshauser.
\newblock Distribution network prices and solar {PV}: {Resolving} rate
  instability and wealth transfers through demand tariffs.
\newblock \emph{Energy Economics}, 54:\penalty0 108--122, 2016.
\newblock \doi{10.1016/j.eneco.2015.11.011}.

\bibitem[Solano et~al.(2018)Solano, Brito, and
  Caamaño-Martín]{solano_impact_2018}
Juan~Carlos Solano, Miguel~C. Brito, and Estefanía Caamaño-Martín.
\newblock Impact of fixed charges on the viability of self-consumption
  photovoltaics.
\newblock \emph{Energy Policy}, 122:\penalty0 322--331, 2018.
\newblock \doi{10.1016/j.enpol.2018.07.059}.

\bibitem[Staffell and Pfenninger(2016)]{staffell_using_2016}
Iain Staffell and Stefan Pfenninger.
\newblock Using bias-corrected reanalysis to simulate current and future wind
  power output.
\newblock \emph{Energy}, 114:\penalty0 1224--1239, 2016.
\newblock \doi{10.1016/j.energy.2016.08.068}.

\bibitem[Tervo et~al.(2018)Tervo, Agbim, DeAngelis, Hernandez, Kim, and
  Odukomaiya]{tervo_economic_2018}
Eric Tervo, Kenechi Agbim, Freddy DeAngelis, Jeffrey Hernandez, Hye~Kyung Kim,
  and Adewale Odukomaiya.
\newblock An economic analysis of residential photovoltaic systems with lithium
  ion battery storage in the {United} {States}.
\newblock \emph{Renewable and Sustainable Energy Reviews}, 94:\penalty0
  1057--1066, 2018.
\newblock \doi{10.1016/j.rser.2018.06.055}.

\bibitem[Thomsen and Weber(2019)]{thomsen_jessica_how_2019}
Jessica Thomsen and Christoph Weber.
\newblock {How the design of retail prices, network charges, and levies affects
  profitability and operation of small-scale PV-Battery Storage Systems}.
\newblock \emph{University of Duisburg-Essen, EWL Working Papers}, 1903, 2019.
\newblock URL \url{https://ideas.repec.org/p/dui/wpaper/1903.html}.

\bibitem[{Vilaça Gomes} et~al.(2018){Vilaça Gomes}, {Knak Neto}, Carvalho,
  Sumaili, Saraiva, Dias, Miranda, and Souza]{gomes_technical-economic_2018}
Phillipe {Vilaça Gomes}, Nelson {Knak Neto}, Leonel Carvalho, Jean Sumaili,
  Joao~T. Saraiva, Bruno~H. Dias, Vladimiro Miranda, and S.~M. Souza.
\newblock Technical-economic analysis for the integration of {PV} systems in
  {Brazil} considering policy and regulatory issues.
\newblock \emph{Energy Policy}, 115:\penalty0 199--206, 2018.
\newblock \doi{10.1016/j.enpol.2018.01.014}.

\bibitem[von Hirschhausen(2017)]{hirschhausen_2017}
Christian von Hirschhausen.
\newblock Prosumage and the future regulation of utilities: An introduction.
\newblock \emph{Economics of Energy \& Environmental Policy}, 6:\penalty0 1 --
  5, 2017.
\newblock \doi{10.5547/2160-5890.6.1.cvh}.

\bibitem[Weniger et~al.(2014)Weniger, Tjaden, and
  Quaschning]{weniger_sizing_2014}
Johannes Weniger, Tjarko Tjaden, and Volker Quaschning.
\newblock Sizing of residential {PV} battery systems.
\newblock \emph{Energy Procedia}, 46:\penalty0 78--87, 2014.
\newblock \doi{10.1016/j.egypro.2014.01.160}.

\bibitem[Wiese et~al.(2019)Wiese, Schlecht, Bunke, Gerbaulet, Hirth, Jahn,
  Kunz, Lorenz, Mühlenpfordt, Reimann, and Schill]{wiese_2019}
Frauke Wiese, Ingmar Schlecht, Wolf-Dieter Bunke, Clemens Gerbaulet, Lion
  Hirth, Martin Jahn, Friedrich Kunz, Casimir Lorenz, Jonathan Mühlenpfordt,
  Juliane Reimann, and Wolf-Peter Schill.
\newblock Open power system data – frictionless data for electricity system
  modelling.
\newblock \emph{Applied Energy}, 236:\penalty0 401 -- 409, 2019.
\newblock \doi{10.1016/j.apenergy.2018.11.097}.

\bibitem[Young et~al.(2019)Young, Bruce, and {McGill}]{young_2019}
Sharon Young, Anna Bruce, and Iain {McGill}.
\newblock Potential impacts of residential {PV} and battery storage on
  {Australia's} electricity networks under different tariffs.
\newblock \emph{Energy Policy}, 128:\penalty0 616--627, 2019.
\newblock \doi{10.1016/j.enpol.2019.01.005}.

\bibitem[Yu(2018)]{yu_prospective_2018}
Hyun Jin~Julie Yu.
\newblock A prospective economic assessment of residential {PV}
  self-consumption with batteries and its systemic effects: {The} {French} case
  in 2030.
\newblock \emph{Energy Policy}, 113:\penalty0 673--687, 2018.
\newblock \doi{10.1016/j.enpol.2017.11.005}.

\bibitem[Zerrahn and Schill(2017)]{dieter_2017}
Alexander Zerrahn and Wolf-Peter Schill.
\newblock Long-run power storage requirements for high shares of renewables:
  review and a new model.
\newblock \emph{Renewable and Sustainable Energy Reviews}, 79:\penalty0
  1518--1534, 2017.
\newblock \doi{10.1016/j.rser.2016.11.098}.

\end{thebibliography}
		
		
\newpage
\appendix 
\renewcommand{\baselinestretch}{1.0}

\begin{appendix}

\renewcommand{\thesection}{A.\arabic{section}}
\addtocounter{footnote}{7}
\section*{Appendix}


\section{Prosumage market situation in Germany}\label{app:sec:germany}
This appendix section provides more details on the market situation for prosumage in Germany. In 2017, $1.6$~million PV plants with a total capacity of~$43$~GW supplied roughly~$7$\% of the national electricity demand~\citep{bsw_solar_statistische_2018}. Small-scale rooftop systems with a capacity of up to~$10$~kW, which are typical for the residential sector, accounted for a share of~$20$\% of national PV capacity~\citep{fraunhofer_ise_photovoltaics_2019}. Still, self-generation and~-consumption of electricity is a niche phenomenon, with only~$6$\% of electricity generated by all PV plants consumed on-site in 2018~\citepalias{eeg_in_zahlen_2018}. 

Yet this number is likely to increase with the ongoing dissemination of battery storage. More than~$100,000$ so-called ``solar battery storage systems'' have been installed in Germany until 2018~\citep{bsw_solar_meilenstein_2018}. The mean price of lithium-ion storage systems had halved between~2013 and~2017 \citep{figgener_wissenschaftliches_2018}, and further cost decreases are likely~\citep{schmidt_future_2017}. Beyond that, the German state-owned development bank ran a so-called market incentive program that subsidized the installation of battery storage connected to small scale PV systems. Also, an increasing number of PV systems will drop out of the feed-in tariff (FIT) scheme after the 20-year supporting period. This will be the case for almost half a million small-scale PV plants by 2030~\citep{opsd_open_2018, wiese_2019}. A great share of this capacity can be expected to be still operational~\citep{schill_prosumage_2017} and could be upgraded with a battery storage to engage in prosumage~\citep{dihk_merkblatt_2018}. 


\section{Model details}\label{app:sec:model}
This appendix section provides more details on the numerical model.


\subsection{Technical model details}\label{app:subsec:model_sets} 
The following tables collect the sets, variables, and parameters contained in the model formulation.

\begin{longtable}{lll}
\caption{Sets included in the model} \\
\label{tab:app:sets} \\
\midrule
Set & Elements & Description \\ 
\toprule
\endfirsthead
\multicolumn{3}{c}{\textit{Continued from previous page}} \\	
\toprule	\\
Set & Elements & Description \\ 
\midrule 
\endhead
\multicolumn{3}{c}{\textit{Continued on next page}}	\\
\endfoot
\endlastfoot
$\mathfrak{C}$ & $\textstyle{con \in }$ $\text{\ \small \{lignite, hardcoal, ccgt, ocgt, bio, oil\}}$ & Conventional generation technologies \\
& & and biomass \\
$\mathfrak{RE}$ & $\textstyle{res\in} \text{\ \small \{onshore wind, offshore wind, pv, run-of-river\}}$ & Renewable generation technologies \\
$\mathfrak{S}$ & $\textstyle{sto\in} \text{ \ \small \{lithium-ion, pumped hydro\}}$ & Storage technologies  \\
$\mathfrak{H}$ & $\textstyle{h\in} \text{\ \small \{1,2,..., 8760 \}}$ & Hours of the year \\
\midrule 
\end{longtable}

\vspace{1cm}

\begin{longtable}{lll}
\caption{Variables included in the model} \\
\label{tab:app:variables} \\
\midrule
Variable & Unit & Description \\ 
\toprule
\endfirsthead
\multicolumn{3}{c}{\textit{Continued from previous page}} \\	
\toprule	\\
Variable & Unit & Description \\ 
\midrule 
\endhead
\multicolumn{3}{c}{\textit{Continued on next page}}	\\
\endfoot
\endlastfoot
\\ [-1ex]
\multicolumn{3}{l}{Prosumage segment} \\
\midrule
$ CU^{pro}_h$ &  MW &  Generation of prosumage household curtailed in hour~$h$ \\
$ E^{m2pro}_h$ &  MW & Energy from grid consumed by prosumage household in hour~$h$ \\
$ G^{pro2m}_h$ &  MW & Generation of prosumage household sent to grid in hour~$h$  \\  
$ G^{pro2pro}_h$ &  MW & Generation of prosumage household directly consumed in hour~$h$
\\ 
$ N^{pro}_{pv}$ &  MW & Installed capacity prosumage household PV system \\
$ N^{pro,E}_{sto}$ &  MWh & Installed capacity prosumage household storage energy\\
$ N^{pro,P}_{sto}$ &  MW & Installed capacity prosumage household storage power\\
$ STO^{in,pro}_h$ &  MW & Storage inflow of prosumage household storage in hour~$h$ \\
$ STO^{l,pro2pro}_h$ &  MWh & Storage level of prosumage household storage in hour~$h$ \\
$ STO^{out,pro}_h$ &  MW & Storage outflow from prosumage household storage in hour~$h$ \\
$ \lambda^{enbal,pro}_{h}$ &  & Dual variable on household energy balance in hour~$h$ \\
$ \lambda^{pv,pro}_{h}$ &  & Dual variable on hourly use of PV energy in hour~$h$ \\
$ \lambda^{pvmax,pro}$ &  & Dual variable on maximum PV investment \\
$ \lambda^{sto,pro}_{h}$ &  & Dual variable on storage level in hour~$h$ \\
$ \lambda^{stoin,pro}_{h}$ &  & Dual variable on maximum storage loading in hour~$h$ \\
$ \lambda^{stol,pro}_{h}$ &  & Dual variable on maximum storage level in hour~$h$ \\
$ \lambda^{stoout,pro}_{h}$ &  & Dual variable on maximum storage in hour~$h$ discharging \\
\\ [-1ex]
\multicolumn{3}{l}{Power sector} \\
\midrule
$ CU_{res,h}$ &  MW & Curtailment renewable technology $res$ in hour~$h$ \\
$ G_{con,h}$ &  MW &  Generation level conventional technology~$con$ in hour~$h$ \\
$ G_{res,h}$ &  MW & Generation renewable technology $res$ in hour~$h$ \\
$ STO^{in}_{sto,h}$ &  MW & Storage inflow technology $sto$ in hour~$h$ \\
$ STO^{l}_{sto,h}$ &  MWh & Storage level technology $sto$ in hour~$h$ \\
$ STO^{out}_{sto,h}$ &  MW & Storage outflow technology $sto$ in hour~$h$ \\
$ \lambda^{con}_{con,h}$ &  & Dual variable on maximum conventional generation in hour~$h$ \\
$ \lambda^{enbal}_{h}$ &  & Dual variable on power sector energy balance in hour~$h$ \\
$ \lambda^{resgen}_{res,h}$ &  & Dual variable on renewable generation in hour~$h$ \\
$ \lambda^{sto}_{sto,h}$ &  & Dual variable on storage level in hour~$h$ \\
$ \lambda^{stoin}_{sto,h}$ &  & Dual variable on maximum storage loading in hour~$h$ \\
$ \lambda^{stol}_{sto,h}$ &  & Dual variable on maximum storage level in hour~$h$ \\
$ \lambda^{stoout}_{sto,h}$ &  & Dual variable on maximum storage discharging in hour~$h$ \\ \\
\midrule
\end{longtable}

\vspace{1cm}

\begin{longtable}{ll}
\caption{Parameters included in the model} \\
\label{tab:app:parameters} \\
\midrule
Parameter  & Description \\ 
\toprule
\endfirsthead
\multicolumn{2}{c}{\textit{Continued from previous page}} \\	
\toprule	\\
Parameter  & Description \\ 
\midrule 
\endhead
\multicolumn{2}{c}{\textit{Continued on next page}}	\\
\endfoot
\endlastfoot
$c^{fix}$  & Annual fixed costs  \\
$c^{inv}$  & Annualized specific investment costs \\
$c^{inv,E}_{sto}$  &  Annualized specific investment costs into storage energy \\
$c^{inv,P}_{sto}$  &  Annualized specific investment costs into storage power \\
$c^{m}$  & Marginal costs (short-term variable costs) \\
$d_{h}$  & Hourly wholesale demand in hour~$h$ \\
$d_{h}^{pro}$  & Hourly demand prosumage household in hour~$h$ \\
$m^{i,max}_{pv}$  & Maximum installable PV capacity for a prosumage household \\
$n_{con}$  & Installed capacity conventional and biomass technologies at the power sector level  \\
$n_{res}$   & Installed capacity renewable technology at the power sector level  \\
$n^{E}_{sto}$   & Installed capacity storage energy at the power sector level  \\
$n^{P}_{sto}$  &  Installed capacity storage energy at the power sector level \\
$t^{ener}_h$  & Energy-related price component of household electricity retail tariff in hour~$h$ \\
$t^{fix}$  &  Fixed annual electricity charge for households in hour~$h$ \\
$t^{other}$  &  Non-energy price component of household electricity retail tariff  \\
$t^{prod}_h$  &  Remuneration for household renewable generation in hour~$h$ \\
$\eta_{sto}$  & Storage round-trip efficiency \\
$\phi_{h}^{avail}$  & Hourly available energy from renewables as fraction of installed capacity  in hour~$h$ \\
\midrule 
\end{longtable} 

\clearpage
\newpage
\pagebreak

\subsection{Prosumage household optimization}\label{app:subsec:model_lagrange_hh}
\renewcommand{\theequation}{A.\arabic{equation}}

The cost minimization problem of the prosumage household is given by equations~\eqref{eq:household_problem}. The corresponding Lagrangian is:
\vspace{0.2cm}
\begin{align}
\mathcal{L} =  
& \hspace{0.2cm} \sum\limits_{h}\left[E^{m2pro}_h \left(t^{ener}_h + t^{other}\right)\right] + t^{fix} \label{eq:obj}  \nonumber \\
- &\sum\limits_{h}\left(G^{pro2m}_h t^{prod}_h\right) \nonumber \\
+  &N_{pv}^{pro} \left(c^{inv}_{pv} + c^{fix}_{pv}\right) + N_{sto}^{pro,E} \left(c^{inv,E}_{sto} + \textstyle\frac{1}{2} c^{fix}_{sto}\right) + N_{sto}^{pro,P} \left(c^{inv,P}_{sto} + \textstyle\frac{1}{2} c^{fix}_{sto}\right) \nonumber \\ \nonumber
+ & \sum\limits_{h} \lambda^{enbal,pro}_{h}\left(d^{pro}_h - G^{pro2pro}_h - STO^{out,pro}_h - E^{m2pro}_h\right)\\ \nonumber
+  &\sum\limits_{h} \lambda^{pv,pro}_{h}\left(G^{pro2pro}_h + G^{pro2m}_h  + CU^{pro}_h + STO^{in,pro}_h - \phi^{avail}_{pv,h}*N_{pv}^{pro}\right)\\ \nonumber
+  &\sum_{h>h_1} \lambda^{sto,pro}_{h}\left(STO^{l,pro}_h - STO^{l,pro}_{h-1} 
- \textstyle\frac{1+ \eta_{sto}}{2}STO^{in,pro}_h \nonumber
+ \textstyle\frac{2}{1+ \eta_{sto}}STO^{out,pro}_h
\right)\\ \nonumber  
+ & \lambda^{sto,pro}_{h_1}\left(STO^{l,pro}_{h_1}
- \textstyle\frac{1+ \eta_{sto}}{2}STO^{in,pro}_{h_1} \nonumber
+ \textstyle\frac{2}{1+ \eta_{sto}}STO^{out,pro}_{h_1}
\right)\\ \nonumber  
+  &\sum_{h}\lambda^{stol,pro}_{h}\left(STO^{l,pro}_h - N_{sto}^{pro,E}\right)\\ \nonumber
+  &\sum_{h}\lambda^{stoin,pro}_{h}\left(STO^{in,pro}_h - N_{sto}^{pro,P}\right)\\\nonumber
+  &\sum_{h}\lambda^{stoout,pro}_{h}\left(STO^{out,pro}_h - N_{sto}^{pro,P}\right)\\
+  &\lambda^{pvmax,pro}\left(N_{pv}^{pro} - m_{pv}\right)\\ \nonumber
\end{align}

The first order KKT conditions corresponding to the household's problem are:
\vspace{0.2cm}
\begin{subequations}\label{eq:kkt_household}
\begin{align}
0 &\leq - t^{prod}_h + \lambda^{pv,pro}_{h} &&\perp G^{pro2m}_h &&&\geq 0 &&&\forall h  
\\ 
0 &\leq - \lambda^{enbal,pro}_{h} + \lambda^{pv,pro}_{h} &&\perp G^{pro2pro}_h &&&\geq 0  &&&\forall h  
\\ 
0 &\leq  t^{ener}_h + t^{other} - \lambda^{enbal,pro}_{h} &&\perp E^{m2pro}_h &&&\geq 0  &&&\forall h 
\\ 
0 &\leq \lambda^{pv,pro}_{h} &&\perp CU^{pro}_h &&&\geq 0  &&&\forall h \\ 
0 &\leq  \lambda^{pv,pro}_{h} - \textstyle\frac{1+ \eta_{sto}}{2}*\lambda^{sto,pro}_{h} + \lambda^{stoin,pro}_{h} &&\perp STO^{in,pro}_h &&&\geq 0  &&&\forall h 
\\ 
0 &\leq - \lambda^{enbal,pro}_{h} + \textstyle\frac{2}{1+ \eta_{sto}}*\lambda^{sto,pro}_{h} + \lambda^{stoout,pro}_{h} &&\perp STO^{out,pro}_h &&&\geq 0  &&&\forall h 
\\
0 &\leq  \lambda^{stol,pro}_{h}  + \lambda^{sto,pro}_{h} - \lambda^{sto,pro}_{h+1} &&\perp STO^{l,pro}_h &&&\geq 0  &&&\forall h>h_1
\\ 
0 &\leq  \lambda^{stol,pro}_{h_1}  + \lambda^{sto,pro}_{h_1} &&\perp STO^{l,pro}_{h_1} &&&\geq 0  &&&
\\ 
0 &\leq c^{inv,E}_{sto} + \textstyle\frac{1}{2} c^{fix}_{sto} - \textstyle\sum_{h}\lambda^{stol,pro}_h  
&&\perp N_{sto}^{pro,E} &&&\geq 0  &&&
\\ 
0 &\leq c^{inv,P}_{sto} + \textstyle\frac{1}{2} c^{fix}_{sto} - \textstyle\sum_{h}(\lambda^{stoin,pro}_h + \lambda^{stoout,pro}_h) &&\perp N_{sto}^{pro,P} &&&\geq 0  &&&   
\\ 
0 &\leq  c^{inv}_{pv} + c^{fix}_{pv} - \textstyle\sum_{h}(\lambda^{pv,pro}_{h}*\phi^{avail}_{pv,h}) +\lambda^{pvmax,pro} &&\perp N_{pv}^{pro} &&&\geq 0  &&& 
\\ 
0 &\leq N_{sto}^{pro,E} - STO^{l,pro}_h &&\perp \lambda^{stol,pro}_{h} &&&\geq 0  &&& \forall h  
\\ 
0 &\leq N_{sto}^{pro,P} - STO^{in,pro}_h  &&\perp \lambda^{stoin,pro}_{h} &&&\geq 0  &&& \forall h  
\\
0 &\leq N_{sto}^{pro,P} - STO^{out,pro}_h &&\perp \lambda^{stoout,pro}_{h} &&&\geq 0  &&&\forall h   
\\
0 &\leq m_{pv} - N_{pv}^{pro} &&\perp \lambda^{pvmax,pro} &&&\geq 0  &&&
\end{align}
\begin{align}
0 &\leq G^{pro2pro}_h + STO^{out,pro}_h + E^{m2pro}_h - d^{pro}_h && , \lambda^{enbal,pro}_{h} \mbox{ free} && \forall h 
\\
0 &\leq \phi^{avail}_{pv,h}* N_{pv}^{pro} - G^{pro2pro}_h - G^{pro2m}_h - CU^{pro}_h \nonumber 
\\
&  \hspace{2.6cm} - STO^{in,pro}_h  && , \lambda^{pv,pro}_{h} \mbox{ free} &&  \forall h 
\end{align}
\begin{align}
0 & \leq  STO^{l,pro}_{h-1} \nonumber
+ \textstyle\frac{1+ \eta_{sto}}{2}STO^{in,pro}_h  
\\ 
&  \hspace{2.cm} - \textstyle\frac{2}{1+ \eta_{sto}}STO^{out,pro}_h - STO^{l,pro}_h  &&  , \lambda^{sto,pro}_{h} \mbox{ free} && \forall {h>h_1} 
\\  \nonumber \\ 
0 & \leq  \textstyle\frac{1+ \eta_{sto}}{2}STO^{in,pro}_{h_1}  - \textstyle\frac{2}{1+ \eta_{sto}}STO^{out,pro}_{h_1}  - STO^{l,pro}_{h_1}  &&  , \lambda^{sto,pro}_{h_1} \mbox{ free} &&\\
\nonumber
\end{align}
\end{subequations}

\pagebreak

\subsection{Power sector dispatch optimization} \label{app:subsec:model_lagrange_sys}
The generation side of the power sector is represented by a benevolent system operator who minimizes total system costs~$Z^{sys}$ through optimal dispatch. This dispatch is equivalent to a competitive equilibrium outcome and results in dispatching generators and storage based on the short-term variable costs or opportunity costs, respectively. The system costs for power generation comprise the short-term variable costs of conventional generators~$c_{con}^m$ only, as shown in the objective function \ref{eq:obj_sys_org}. Renewable generation technologies and storage are assumed to incur no variable operating costs.
\vspace{0.2cm}
\begin{subequations}\label{eq:optimization_system}
\begin{align}
\mbox{ min} \hspace{0.1cm} Z^{sys} = &\sum\limits_{h}\sum\limits_{con} c_{con}^m G_{con,h} \label{eq:obj_sys_org}
\end{align}
    
Equation~\ref{eq:ener_balancesys} shows the energy balance, i.e., the market clearing condition, which must hold in each hour. The sum of non-prosumage consumer demand~$d_h$, prosumage grid energy demand~$E^{m2pro}_h$, and storage intake~$STO_{sto,h}^{in}$ must equal generation of conventional power plants~$G_{con,h}$, renewable generation~$G_{res,h}$, storage discharging~$STO^{out}_{sto,h}$, and prosumage energy fed to the grid~$ G^{pro2m}_h$. The dual variable to this constraint~$\lambda^{enbal}_{h}$ is interpreted as the wholesale electricity price. It indicates the marginal costs associated with a marginal increase on the demand-side.
\vspace{0.2cm}
\begin{align} \nonumber
d_h + \sum\limits_{sto}STO_{sto,h}^{in} + E^{m2pro}_h &= \sum\limits_{con} G_{con,h} + \sum\limits_{res} G_{res,h} + G^{pro2m}_h   \\
& \hspace{2.2cm}  + \sum\limits_{sto} STO^{out}_{sto,h}
 && \forall h\label{eq:ener_balancesys} &&& (\lambda^{enbal}_{h})
\end{align} 

As is the case for prosumage households, energy generation of each renewable technology~$res$ at the power sector level depends on the hourly capacity factor~$\phi^{avail}_{res,h}$ and the exogenous capacity~$n_{res}$. Equation~\ref{eq:res_use} prescribes that hourly available renewable energy can either be curtailed~$CU_{res,h}$ or consumed on the market~$G_{res,h}$:
\vspace{0.2cm}
\begin{align}
 \phi^{avail}_{res,h}*n_{res} &= G_{res,h} + CU_{res,h} && \forall res, h\label{eq:res_use} &&& (\lambda^{resgen}_{res,h})
\end{align}

Conventional generation is perfectly dispatchable and only constrained by the installed capacity~$n_{con}$ as shown in equation~\ref{eq:con_l2_sys}. 
\vspace{0.2cm}
\begin{align}
G_{con,h} &\leq n_{con}   \label{eq:con_l2_sys}  &&\forall con, h &&&(\lambda^{con}_{con,h}) 
\end{align}

Besides conventional and renewable generation technologies, pumped-hydro power storage is also included for power supply on the power sector level. Equation~\ref{eq:sto_l_sys} describes the change of the energy level of the storage over time. It is the sum of the energy level of the prior period~$STO^{l}_{sto,h-1}$ and current storage charging~$STO^{in}_{sto,h}$, minus the energy discharged~$STO^{out}_{sto,h}$. Both charging and discharging activities account for storage energy losses.
\vspace{0.2cm}
\begin{align}
STO^{l}_{sto,h} &=  STO^{l}_{sto,h-1} + \textstyle\frac{1+\eta_{sto}^{}}{2}STO^{in}_{sto,h} \label{eq:sto_l_sys} \nonumber \\  
& {\hspace{2.4cm}} - \textstyle\frac{2}{1+\eta_{sto}^{}}STO^{out}_{sto,h} &&\forall sto, h>h_1  &&&(\lambda^{sto}_{sto,h}) \\ \nonumber \\
STO^{l}_{sto,h_1} &= \textstyle\frac{1+\eta_{sto}^{}}{2}STO^{in}_{sto,{h_1}} \label{eq:sto_l_sys1} 
- \frac{2}{1+\eta_{sto}^{}}STO^{out}_{sto,{h_1}} &&\forall sto &&&(\lambda^{sto}_{sto,h_1})
\end{align}

Furthermore, the storage energy level is constrained by its energy capacity as given by equation~\ref{eq:sto_l_cap}. Likewise, storage charging and discharging cannot exceed the installed power capacity.
\vspace{0.2cm}
\begin{align}
\label{eq:sto_l_cap} STO^{l}_{sto,h} &\leq n_{sto}^{E} && \forall sto, h &&& (\lambda^{stol}_{sto,h}) \\
\label{eq:sto_in_cap} STO^{in}_{sto,h}  &\leq n_{sto}^{P} &&\forall sto, h  &&& (\lambda^{stoin}_{sto,h})  \\
\label{eq:sto_out_cap} STO^{out}_{sto,h} &\leq n_{sto}^{P} &&\forall sto, h  &&& (\lambda^{stoout}_{sto,h}) 
\end{align}

The Lagrangian corresponding to the power sector dispatch problem is:
\vspace{0.2cm}
\begin{align}
\mathcal{L} =  
&\sum\limits_{h}\sum\limits_{con} \left(c_{con}^m G_{con,h}\right) \nonumber \\\label{eq:obj_sys}  \nonumber
    + &\sum\limits_{h} \lambda^{enbal}_{h}\left(d_h + \sum\limits_{sto}STO_{sto,h}^{in}+ E^{m2pro}_h - \sum\limits_{con} G_{con,h} - \sum\limits_{res} G_{res,h} - \sum\limits_{sto} STO^{out}_{sto,h} - G^{pro2m}_h\right)\nonumber \\\nonumber
 + &\sum\limits_{h}\sum\limits_{res}\lambda^{resgen}_{res,h}\left( G_{res,h} + CU_{res,h} - \phi^{avail}_{res,h}n_{res}\right)  \\\nonumber
+ &\sum\limits_{h>h_1} \sum\limits_{sto} \lambda^{sto}_{sto,h} \left(STO^{l}_{sto,h} - STO^{l}_{sto,h-1} - \textstyle\frac{1+\eta_{sto}}{2}STO^{in}_{sto,h}
 + \textstyle\frac{2}{1+\eta_{sto}}STO^{out}_{sto,h}\right) \\ \nonumber
 + &\sum\limits_{sto}\lambda^{sto}_{sto,h_1} \left(STO^{l}_{sto,h_1} - \textstyle\frac{1+\eta_{sto}}{2}STO^{in}_{sto,{h_1}} 
 + \frac{2}{1+\eta_{sto}}STO^{out}_{sto,{h_1}}\right) \\\nonumber
+ &\sum\limits_{h}\sum\limits_{con}\lambda^{con}_{con,h} \left( G_{con,h} - n_{con}\right)\\ \nonumber 
+ &\sum\limits_{h}\sum\limits_{sto} \lambda^{stol}_{sto,h}\left(STO^{l}_{sto,h} - n_{sto}^{E} \right) \\ \nonumber 
+ &\sum\limits_{h}\sum\limits_{sto}\lambda^{stoin}_{sto,h} \left(  STO^{in}_{sto,h} - n_{sto}^{P} \right) \\  
+ &\sum\limits_{h}\sum\limits_{sto}\lambda^{stoout}_{sto,h}\left( STO^{out}_{sto,h} - n_{sto}^{P} \right) \\ \nonumber 
\end{align}
\end{subequations}

The first order KKT conditions corresponding to the power sector dispatch are given by:
\vspace{0.2cm}
\begin{subequations}\label{eq:kkt_system}
\begin{align}
 0 &\leq - \lambda^{enbal}_{h} + \lambda^{resgen}_{res,h}
 &&\perp G^{}_{res,h} &&&\geq 0 &&&\forall res, h \\ 
0 &\leq \lambda^{stol}_{sto,h} + \lambda^{sto}_{sto,h} - \lambda^{sto}_{sto,h+1}  &&\perp STO^{l}_{sto,h} &&&\geq 0 &&&\forall sto, {h>h_1}  \\ 
0 &\leq \lambda^{stol}_{sto,h_1} + \lambda^{sto}_{sto,h_1} &&\perp STO^{l}_{sto,h_1} &&&\geq 0 &&&\forall sto  \\ 
0 &\leq\lambda^{enbal}_{h} - \textstyle\frac{1+\eta_{sto}}{2}\lambda^{sto}_{sto,h} + \lambda^{stoin}_{sto,h}
&&\perp STO^{in}_{sto,h} &&&\geq 0 &&&\forall sto, h \\ 
0 &\leq - \lambda^{enbal}_{h} + \textstyle\frac{2}{1+\eta_{sto}}\lambda^{sto}_{sto,h} + \lambda^{stoout}_{sto,h}
&&\perp STO^{out}_{sto,h} &&&\geq 0 &&&\forall sto, h  
\end{align}
 \begin{align}
0 &\leq n_{con} - G_{con,h} &&\perp \lambda^{con}_{con,h} &&&\geq 0  &&& \forall con, h\\
0 &\leq n_{sto}^{E} - STO^{l}_{sto,h} &&\perp \lambda^{stol}_{sto,h} &&&\geq 0  &&& \forall sto, h \\
0 &\leq n_{sto}^{P} - STO^{in}_{sto,h} &&\perp \lambda^{stoin}_{sto,h} &&&\geq 0  &&& \forall sto, h \\
0 &\leq n_{sto}^{P} - STO^{out}_{sto,h} &&\perp  \lambda^{stoout}_{sto,h} &&&\geq 0  &&& \forall sto, h
\end{align}
\begin{align}
  0 & \leq \textstyle\sum\limits_{con} G_{con,h}\nonumber
 +\label{eq:kktener} \sum\limits_{res} G_{res,h} + \sum\limits_{sto} STO^{out}_{sto,h} + G^{pro2m}_h \\ 
  & \hspace{2.1cm} - d_h -\sum\limits_{sto}STO_{sto,h}^{in} - E^{m2pro}_h
   &&  , \lambda^{enbal}_{h} \mbox{ free} &&& \forall h\\ \nonumber \\
 0 & \leq  \phi^{avail}_{res,h}*n_{res} - G_{res,h} - CU_{res,h} &&  , \lambda^{resgen}_{res,h}\mbox{ free} &&& \forall res, h\\ \nonumber \\
 0 & \leq STO^{l}_{sto,h-1} +  \textstyle\frac{1+\eta_{sto}}{2}STO^{in}_{sto,h} \nonumber \\
  &  \hspace{2.4cm} - \textstyle\frac{2}{1+\eta_{sto}}STO^{out}_{sto,h} - STO^{l}_{sto,h} &&  , \lambda^{sto}_{sto,h}  \mbox{ free} &&& \forall sto, {h>h_1}\\ 
\nonumber \\
 0 &  \leq \textstyle\frac{1+\eta_{sto}}{2}STO^{in}_{sto,h_1}
 - \frac{2}{1+\eta_{sto}}STO^{out}_{sto,h_1} - STO^{l}_{sto,h_1} &&  , \lambda^{sto}_{sto,h_1} \mbox{ free} &&& \forall sto
\end{align}
\end{subequations}

Equations~\ref{eq:kkt_household} and~\ref{eq:kkt_system} are combined to form the MCP that it solved to determine numerical results.

\newpage
\clearpage
\pagebreak

\subsection{Power sector data}\label{app:subsec:model_data}
Table~\ref{tab:app:parameters_system} compiles numerical assumptions on technologies in the central power sector. 

\renewcommand{\baselinestretch}{1.34}

\begin{longtable}{L{4cm}lllL{4.5cm}}
\caption{Technical assumptions on conventional power plants and centralized pumped-hydro storage} \\
\label{tab:app:parameters_system} \\
\midrule
Parameter & Symbol & Value & Unit & Source \\ 
\toprule
\endfirsthead
\multicolumn{5}{c}{\textit{Continued from previous page}} \\	
\toprule	\\
Parameter & Symbol & Value & Unit & Source \\ 
\midrule 
\endhead
\multicolumn{5}{c}{\textit{Continued on next page}}	\\
\endfoot
\endlastfoot
\textbf{Market assumptions} &&&&  \\
CO$_{2}$ price &  & 29.4  & EUR/t & \citet{bnetza_genehmigung_2018} \\
Interest rate & & 4\%  & - & Own assumption \\
\midrule 
\textbf{Lignite} &&&&  \\
Thermal efficiency &  & 0.38  & - & \citet{schroder_current_2013} \\
Carbon content &  & 0.311 &  t CO$_{2}$/MWh & \citet{bnetza_genehmigung_2018}  \\
Fuel price &  & 5.6 & EUR/MWh & \citet{bnetza_genehmigung_2018} \\
Marginal generation costs & $c^{m}_{lignite}$ & 38.8 &  EUR/MWh & Own calculation \\
\midrule 
\textbf{Hard Coal} &&&&  \\
Thermal efficiency &  & 0.43 & - &  \citet{schroder_current_2013} \\
Carbon content &  & 0.26 & t CO$_{2}$/MWh & \citet{bnetza_genehmigung_2018} \\
Fuel price & & 8.4 &  EUR/MWh & \citet{bnetza_genehmigung_2018} \\
Marginal generation costs & $c^{m}_{hardcoal}$  & 37.31 & EUR/MWh & Own calculation \\
\midrule 
\textbf{CCGT} &&&&  \\
Thermal efficiency &  & 0.542 & - &  \citet{schroder_current_2013} \\
Carbon content &  & 0.155 & t CO$_{2}$/MWh & \citet{bnetza_genehmigung_2018} \\
Fuel price &  & 26.4 & EUR/MWh & \citet{bnetza_genehmigung_2018} \\
Marginal generation costs & $c^{m}_{ccgt}$ &  57.12 & EUR/MWh & Own calculation \\
\midrule 
\textbf{OCGT} &&&&  \\
Thermal efficiency &  & 0.4 & -  & \citet{schroder_current_2013} \\
Carbon content &  & 0.155 & t CO$_{2}$/MWh & \citet{bnetza_genehmigung_2018} \\
Fuel price &  & 26.4 & EUR/MWh & \citet{bnetza_genehmigung_2018} \\
Marginal generation costs & $c^{m}_{ocgt}$  & 77.39 & EUR/MWh & Own calculation \\
\midrule
\textbf{Oil} &&&&  \\
Thermal efficiency &  & 0.35 & -  & \citet{schroder_current_2013} \\
Carbon content & & 0.216 & t CO$_{2}$/MWh & \citet{bnetza_genehmigung_2018} \\
Fuel price &  & 48.3 & EUR/MWh & \citet{bnetza_genehmigung_2018} \\
Marginal generation costs & $c^{m}_{oil}$ &  156.14 & EUR/MWh & Own calculation \\
\midrule 
\textbf{Biomass} &&&&  \\
Thermal efficiency &  & 0.487 & -  & \citet{schroder_current_2013} \\
Carbon content  &  & 0.00 & $t CO_{2} MWh_{th}$  & \citet{bnetza_genehmigung_2018} \\
Fuel price &  & 10 & $EUR/MWh_{th}$ & \citet{bnetza_genehmigung_2018} \\
Marginal generation costs & $c^{m}_{bio}$ &  20.53 & EUR/MWh & Own calculation \\
\midrule
\multicolumn{5}{l}{\textbf{Pumped hydro storage}}  \\
Round-trip efficiency & $\eta_{hydro}$ & 0.8 & - & \citet{pape_roadmap_2014} \\
\bottomrule
\end{longtable}

\clearpage
\pagebreak

\end{appendix}


\end{document}